\documentclass[preprint,12pt]{elsarticle}

\usepackage{amssymb}
\usepackage{url}
\usepackage{hyperref}
\usepackage{fullpage}
\usepackage{graphics}
\usepackage{grffile}

\journal{}

\begin{document}

\begin{frontmatter}

\title{Modeling morphology evolution during solvent-based fabrication of organic solar cells}

\author[ME]{Olga Wodo}
\author[ME,ECpE]{Baskar Ganapathysubramanian\corref{cor}}
\address[ME]{Department of Mechanical Engineering, Iowa State University, Ames, Iowa,}
\address[ECpE]{Department of Electrical and Computer Engineering, Iowa State University, Ames, Iowa}
\cortext[cor]{Tel.:+1 515-294-7442; fax: +1 515-294-3261. E-mail address: baskarg@iastate.edu, URL: http://www3.me.iastate.edu/bglab/ }

\begin{abstract}

Solvent-based thin-film deposition constitutes a popular class of fabrication strategies for manufacturing organic electronic devices like organic solar cells. All such solvent-based techniques usually involve preparing dilute blends of electron-donor and electron-acceptor materials dissolved in a volatile solvent. After some form of coating onto a substrate to form a thin film, the solvent evaporates. An initially homogeneous mixture separates into electron-acceptor rich and electron-donor rich regions as the solvent evaporates.
Depending on the specifics of the blend, processing conditions, and substrate characteristics different morphologies are typically formed. Experimental evidence consistently confirms that the resultant morphology critically affects device performance. A computational framework that can predict morphology evolution can significantly augment experimental analysis. Such a framework will also allow high throughput analysis of the large phase space of processing parameters, thus yielding considerable insight into the process-structure-property relationships governing organic solar cell behavior.

In this paper, we formulate a computational framework to predict evolution of morphology during solvent-based fabrication of organic thin films. This is accomplished by developing a phase field-based model of {\it evaporation-induced and substrate-induced phase-separation in ternary systems}. This formulation allows most of the important physical phenomena affecting morphology evolution during fabrication to be naturally incorporated. We discuss the various numerical and computational challenges associated with a three dimensional, finite-element based, massively parallel implementation of this framework. This formulation allows, for the first time, to model three-dimensional nanomorphology evolution over large time spans on device scale domains.
We illustrate this framework by investigating and quantifying the effect of various process and system variables on morphology evolution. We explore ways to control the morphology evolution by investigating different evaporation rates, blend ratios and interaction parameters between components.
\end{abstract}


\end{frontmatter}

\section{Introduction}
\label{ch:intro}

Organic solar cells (OSC) manufactured from organic blends~\cite{DSB09,HS04} represent a promising low-cost, rapidly deployable strategy for harnessing solar energy~\cite{PRB09}. 
In contrast to traditional silicon based solar cells, organic (or polymer) solar cells are low weight, printable on flexible substrate, and most importantly, can be produced at room temperature at very low cost. 

\begin{figure}[h]
\centering
\includegraphics[width=0.5\textwidth]{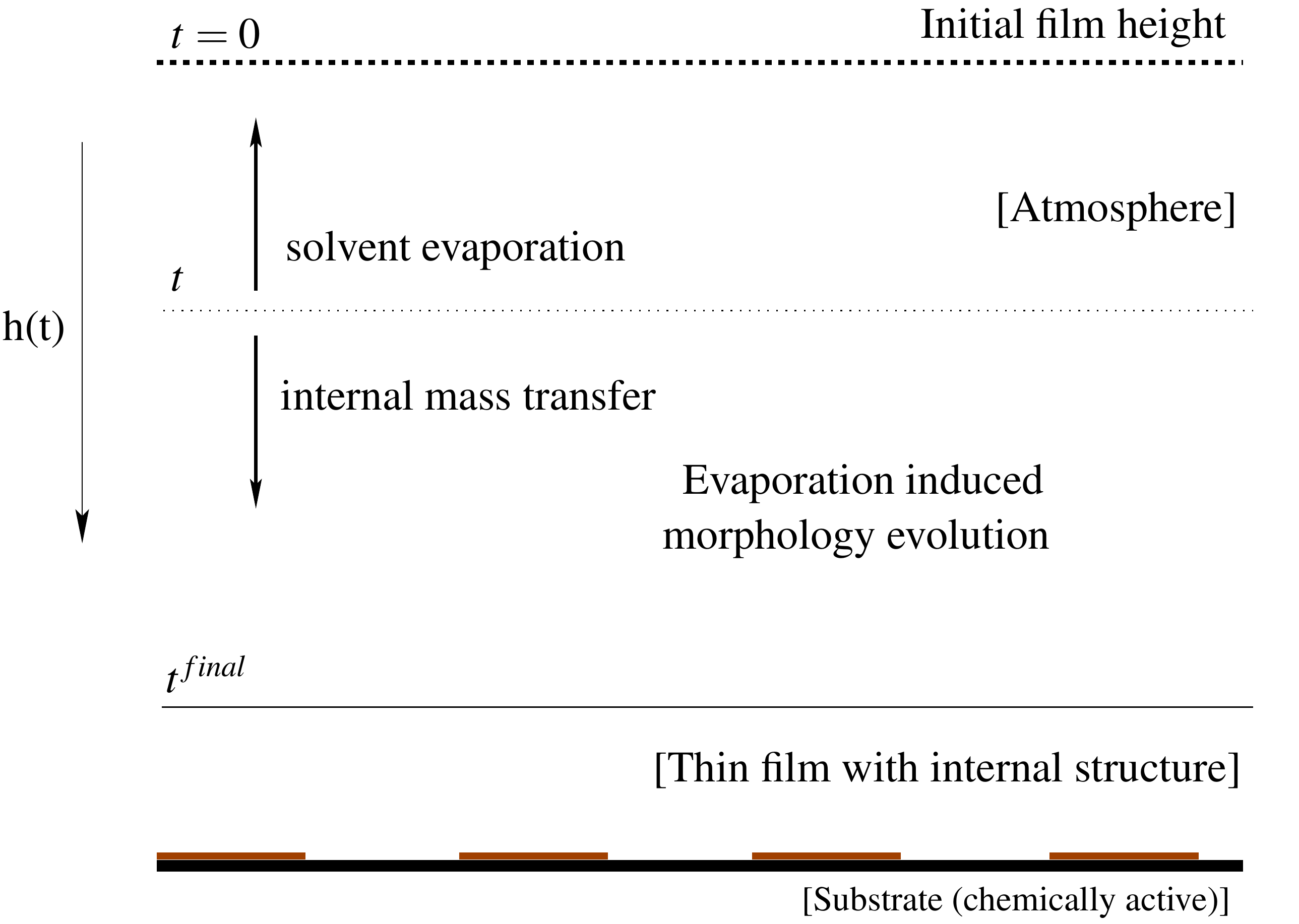}
\includegraphics[width=0.45\textwidth]{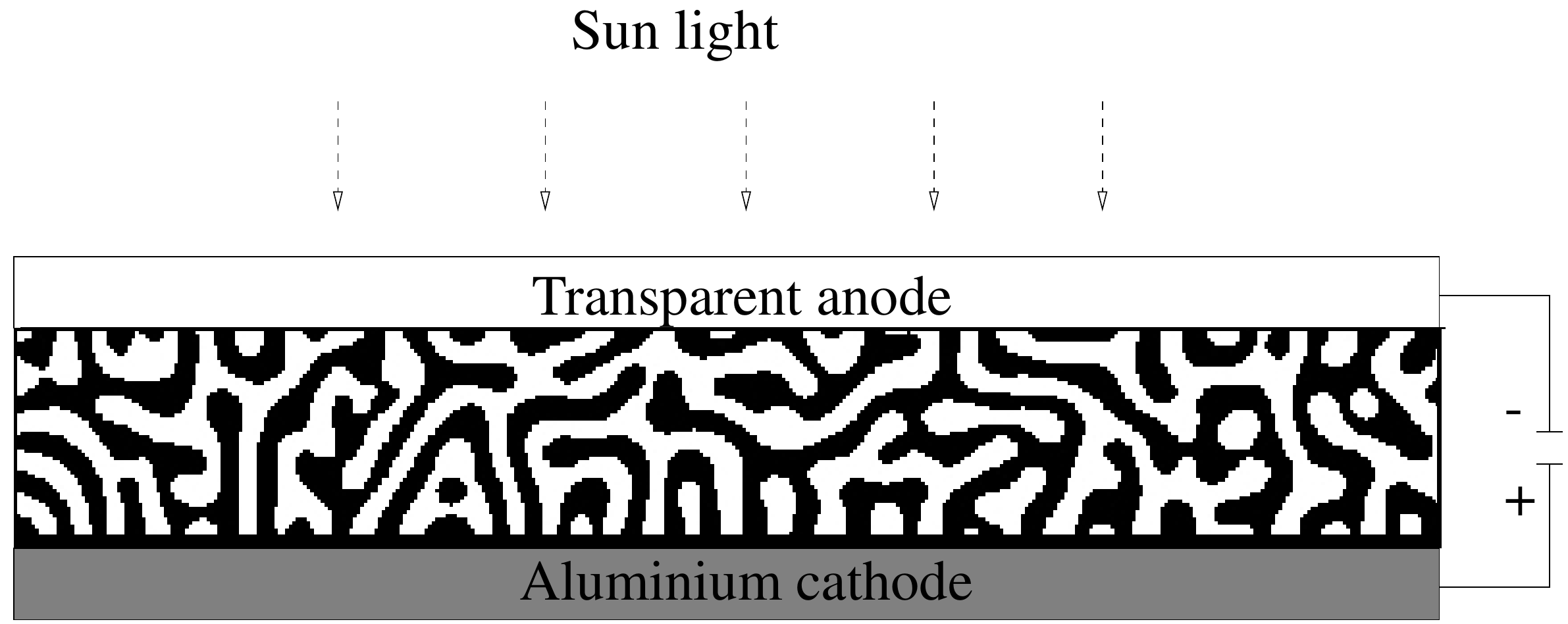}
\caption{Schematic illustration of phenomena during solvent-based fabrication techniques (left). Solution of donor, acceptor and solvent is coated on the chemically active substrate. As the solvent evaporates into the atmosphere, morphology evolution is triggered. (Right) An example thin film sandwiched between two electrodes. Internal structure represents the typical morphology of OSC, i.e. the mixture of electron-donor regions (black) and electron-acceptor regions (white) sandwiched between electrodes.}
\label{fig:evapse:scheme}
\end{figure}

Solvent-based thin-film deposition technologies~\cite{K09} (e.g., spin coating, drop casting, doctor blading, roll-to-roll manufacturing) are the most common organic photovoltaic manufacturing techniques. These techniques, especially doctor blading and roll manufacturing, are very attractive, due to their ease of scale-up for large commercial production. All solution-processing techniques usually involve preparing dilute solutions of electron-donor and electron-accepting materials in a volatile solvent. After some form of coating onto a substrate, the solvent evaporates. An initially homogeneous mixture separates into electron-accepting rich regions and electron-donor rich regions as the solvent evaporates~(see Figure~\ref{fig:evapse:scheme}). Depending on the specifics of the polymer blend and processing conditions (spin coating time~\cite{CQFA08,CYH09}, solvent type~\cite{HS06,KSY09}, nature of substrates), different morphologies are typically formed in the active layer. 

The active layer of OSC is a blend of two types of materials: electron donating and electron accepting material. The active layer is sandwiched between electrodes (see Figure~\ref{fig:evapse:scheme} right).  
The morphological distribution of the electron-donor and electron-acceptor subregions in the device strongly determines the power conversion efficiency (PCE) of OSCs~\cite{HS06,BAZ09}. In fact, every stage of the photovoltaic process is affected by this morphology. 
Consequently, there is immense interest to understand morphology evolution during fabrication~\cite{HS06,CHGY09,ChenRussell2011a}. An important technological goal (that understanding morphology evolution will help achieve) is the ability to design fabrication processes to obtain tailored morphologies for high efficiency OSC devices. 
	
Current state-of-the-art approaches towards tailoring the manufacturing process are limited to combinatorial trial-and-error-based experimental investigations~\cite{HS06,CHGY09,LSH05}. Furthermore, the inability to experimentally visualize morphology evolution hinders the ability to quantify the effect of  various process and system variables (such as evaporation rate, blend ratio, solvent type) on morphology evolution. Experimental techniques provide only limited data for analysis: (a)~mainly limited to observations of the lateral organization of the top layer, and (b)~mainly limited to the final morphology. In addition, visualizing 3D morphology remains challenging.
These challenges serve as a compelling rationale for developing a computational framework that can model morphology evolution, thus significantly augmenting experimental analysis. 

Computational approaches to this problem exist, but are mostly limited to one scale: atomistic scale or macro scale. From a macro scale perspective the problem of thin film formation is well-studied. The series of work summarized in~\cite{NGSL05} link the macroscale film thickening during evaporation process with angular velocity of the coater, concentration, evaporation rate, and solution viscosity. {\it However, morphology evolution is not analyzed in these studies.} On the other end of the spectrum, morphology evolution in a typical OSC system was recently studied using molecular dynamics simulations~\cite{HMF10}. The authors were able to predict phase separation between two typical components in a cubical domain of size $25\;nm$ and only for $135\;ns$, without incorporating the macroscale effects of evaporation or substrate. 
To the authors best knowledge, there exists no meso-scale approach that links morphology evolution at the nano scale with macro-scale phenomena like evaporation, and substrate patterning. The development of such a model will provide significant advantages, it will in particular:
\begin{itemize}\vspace{-0.05in}
   \item Serve as a powerful tool to analyze morphology evolution over time in three dimensions. This can be used as a ``stereological microscope'' to visualize morphology evolution from early stages until the formation of the stable morphology. It is worth mentioning that three dimensional experimental reconstruction of the final morphology is possible using electron tomography~\cite{BAZ09,BSWL09,BSW09} but this approach to polymeric systems is exceedingly rare because of the required proper contrast between components.
   \item Allow for independent control over various process and system variables, thus making it easy to isolate factors affecting the process (such as substrate patterning, solvent annealing, or blend ratio).
   \item Ability to perform high throughput analysis. Such an ability allows automated exploration of the phase space of various manufacturing and system variables (such as different blend ratios, evaporation rates, solvent choices, effect of substrates) to understand their effect on morphology and performance. This opens up the possibility of data-driven knowledge discovery to understand the effects of different competing phenomena and, subsequently, tailoring the manufacturing process.
\end{itemize}

In this work, we develop a computational framework to model morphology evolution during fabrication of organic solar cells. We formulate a model that takes into account all the important processes occurring during solvent-based fabrication of OSCs. The model is based on a phase field approach to describe the behavior of multicomponent system with various driving forces. We develop an efficient numerical framework that enables three dimensional, long time-scale analysis of the fabrication.  We illustrate the framework by investigating the effect of independent control over various external conditions: solvent evaporation, blend ratio, and interaction parameters. We further quantify the interplay between the solvent evaporation and diffusion within the film. To the best knowledge of the authors, this is the first comprehensive effort to construct a virtual framework to study 3D morphology evolution during solvent-based fabrication of organic solar cells. 


\section{Physical phenomena affecting morphology evolution during fabrication}

All solution-processing techniques usually involve preparing dilute solutions of electron-donor and electron-accepting materials in a volatile solvent. After some form of coating onto a substrate, the solvent evaporates. An initially homogeneous mixture separates into electron-accepting rich regions and electron-donor rich regions as the solvent evaporates. Depending on the specifics of the blend and processing conditions (spin coating time~\cite{CQFA08,CYH09}, annealing time~\cite{JKN09,KBN09}, solvent type~\cite{HS06,KSY09}, nature of substrates), different morphologies are typically formed. 
The two materials usually used in fabricating OSC are a conjugated polymer and a fullerene derivative, The conjugated polymer is the electron donor material, while the fullerene derivative is the electron-acceptor. We will use the terms electron-accepting material (electron-donating material), acceptor (donor) and fullerene (polymer) interchangeably in this article. 

There is a rich and complex collection of interacting phenomena that direct the morphology evolution during solvent-based fabrication. {\it Phase-separation}~\cite{CYH09,HovenBazan2010,PeetBazan2009,XuYang2009} is a key phenomena triggered by the {\it evaporation of the volatile solvent}. The atmosphere on the free surface determines the evaporation rate of the solvent. The resulting morphology can form multiple phases, making the system a {\it multi-component and multi-phase} system. In addition, the morphology on the substrate's surface may differ from that in the `bulk' state~\cite{CQFA08,CLK08}. Both {\it free surface and substrate} influence the organization of the morphology and directly affect characteristics of the devices. In particular, chemical and physical patterning of the substrate have been shown to affect morphology evolution. In order to enable predictive modeling, each of the following phenomena must be included in the computational model. 

\subsection{Evaporation induced morphology evolution} 

The rate at which solvent is removed from the top layer depends on various factors (e.g. solvent volatility, spinning velocity during spin-coating).
During evaporation, the initially dilute solution becomes enriched in the two solutes due to depletion of solvent. This enrichment results in increased interaction between the solutes and triggers morphology evolution. The evolution is critically determined by the evaporation profile and diffusion of solutes (and solvent) within the film.

\begin{figure}
\centering
\includegraphics[width=0.4\textwidth]{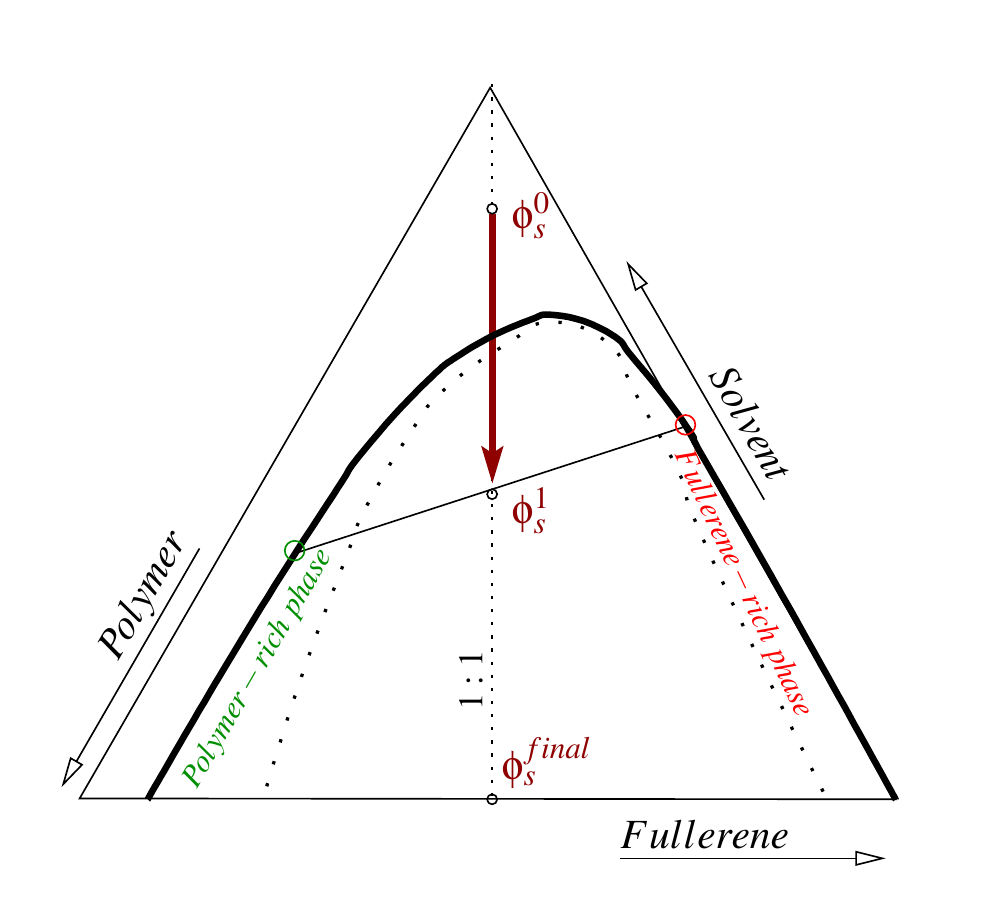}
\includegraphics[width=0.4\textwidth]{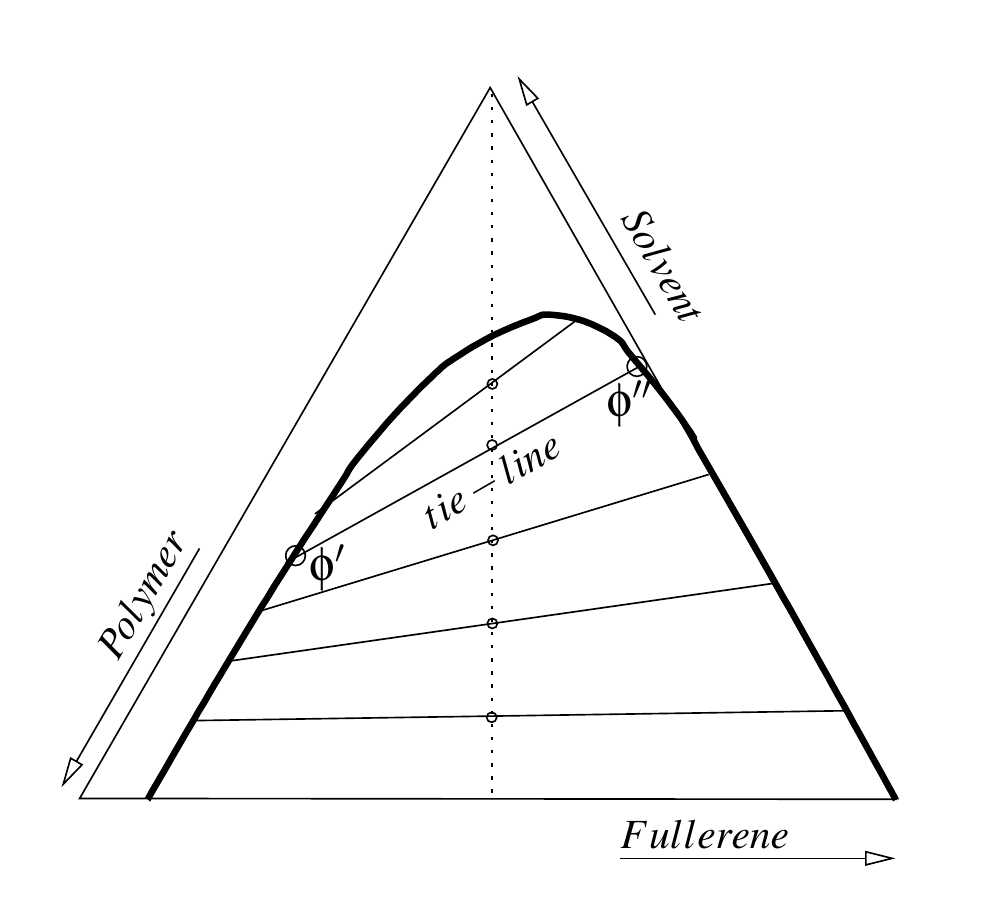}
\caption{Scheme of typical ternary phase diagram. (Left) Typical characteristic curves: binodal line (solid line) and spinodal line (dotted line). 
The spinodal line is related to the position where:  $\frac{\partial^2 f}{\partial \phi_s^2} \frac{\partial^2 f}{\partial \phi_p^2}-\frac{\partial^2 f}{\partial \phi_s\partial \phi_p}=0$.
The binodal curve is related to the equilibrium phase boundary between the single phase and the phase separated region.  
During typical fabrication, initial dilute solution, $\phi_s^0$, is pushed into two-phase region (delimited by the spinodal line) as solvent evaporates ($\phi_s^0 \rightarrow \phi_s^1$). 
Solution separated into polymer-rich ($\phi'$) and fullerene-rich phase ($\phi''$). 
The equilibrium compositions of two phases lie on the binodal curve. 
Line connecting corresponding equilibrium compositions are called tie-lines (right). 
}
\label{fig:PS:scheme}
\end{figure}

As the solvent is removed, the blend is pushed into the spinodal range (immiscible conditions) and induces phase separation (see Figure~\ref{fig:PS:scheme} left). 
Phase separation (or spinodal decomposition) is a mechanism by which solution separates to create phases of different properties.
Under these conditions, the solution is unstable and even small fluctuations lead to fast phase separation. 
The composition of phases changes and reaching equilibrium concentrations. 
Usually, the solution separates into two phases. One phase is rich in donor, the other phase is rich in acceptor material.
The exact composition of phases is determined by thermodynamic conditions~\cite{Robeson07}. 
The creation of the two phases is followed by slow coarsening. The kinetics of phase separation and coarsening is affected by the kinetics of solvent removal.

\subsection{Substrate induced morphology evolution} 
In a confined thin film geometry, the evolution of the morphology close to the walls/interfaces can be significantly different from that of the bulk. In particular, chemical interactions between the substrate and the solute as well as surface patterning can significantly affect morphology evolution. Recent experimental studies of OSC explore this possibility~\cite{CQFA08,ChenYang2010}. By changing substrate properties and by (chemically or physically) patterning the substrate, it was possible to direct vertical segregation, percolation and control phase separation.

In certain systems, crystallization is an additional mechanism of morphology evolution. While incorporating crystallization is relatively straightforward in the current framework, we primarily focus on modeling evaporation and substrate induced spinodal decomposition in this paper. This is in line with experimental results~\cite{NBB07} which seem to suggest that phase separation is a key mechanism in morphology evolution for polymer-polymer blends. 

\section{Phase field approach to model multi-component evolution}
Phase field methods have been used to model morphology evolution in heterogeneous materials typically consisting of grains or domains characterized by different structure, orientation or chemical composition. These methods are highly versatile (due to a diffuse-interface formulation) and easily represent the evolution of complex morphologies with no assumption made about shape or distribution of domains. Various thermodynamic driving forces for morphology evolution, e.g. bulk energy of the system, interfacial energy, substrate energy can be easily introduced. Additionally, the effect of different transport processes, such as mass diffusion, convection, or shear rates, can be directly introduced and exploited using this technique. 


The advantages and flexibility offered by the phase field method provide an ideal framework to model morphology evolution during fabrication of OSC. In particular, the mechanisms that direct morphology evolution (spinodal decomposition and crystallization) can be naturally modeled using this method. They have been well studied for alloy system (dendritic growth and spinodal decomposition in alloy system~\cite{CH58a,KarmaRappel1996}). 
Additionally, various driving forces for phase transformation, such as effect of substrate and evaporation, can be introduced in a very straightforward way without substantial model reformulation. Finally, phase field methods can be scaled to predict morphology evolution for device-scale problems, which is currently impossible using any other framework, like molecular dynamics. 


\section{Mathematical model of evaporation-- and substrate--induced phase separation}

In this section, we formulate the phase field model to simulate morphology evolution in a ternary system with solvent evaporation and substrate interaction included into the model. 

\label{ch:Mo}
\subsection{Ternary phase field model}
Formulating a phase field model usually consists of three stages: (1)~identifying order parameters; (2)~postulating free energy functional that depends on the order parameters; (3)~constructing governing equations that describe the evolution of the system towards a minimum energy state.

\subsubsection{Phase field order parameters}
We consider a ternary system consisting of polymer, fullerene, and solvent. 
We denote the volume fractions of polymer, fullerene and solvent as $\phi_p$, $\phi_f$ and $\phi_s$, respectively. We set the volume fractions $\phi_i$ as the conserved order variables (since $\phi_p+\phi_f+\phi_s=1$). Note that during evaporation, while volume is not conserved, the sum of the volume fractions~-- by definition~-- is conserved.

\subsubsection{Free energy functional}
In the second step, we construct the energy functional for this system. Energy, $F$, consists of two bulk terms: homogeneous energy, and interfacial energy between phases. The total energy is given as:
\begin{equation}
F(\phi_p,\phi_f,\phi_s)=\displaystyle \int_V \left[f(\phi_{p},\phi_f,\phi_s)+\sum_{i=p,f}\displaystyle \frac{\epsilon_{i}^{2}}{2}|\nabla \phi_{i}|^{2} \right] dV+ F_s(\phi_p,\phi_f,\textbf{x})
\label{eq:Energy}
\end{equation}
The energy of the homogeneous system, $f$, also called configurational energy or free energy of mixing, is the quantity which governs phase separation. Homogeneous energy depends only on local volume fractions and is at least double-welled. Wells correspond to the equilibrium concentrations of separated phases.  In contrast, the interfacial energy depends on the composition gradient and is scaled by an interfacial parameter $\epsilon$ (see second term in Eq.~\ref{eq:Energy}). In the current work, we assume\footnote{Interfaces between different phases of different composition have different interfacial energies. However due to limited knowledge about interface properties, this simplification is common~\cite{ZP06,SKP07}.} $\epsilon_p=\epsilon_f=\epsilon$.

We construct homogeneous energy, $f$, using the Flory-Huggins formulation~\cite{Robeson07,JonesRichard99,Strobl07}, which is suitable for polymer solutions.~\footnote{Following~\cite{SKP07}, we include an additional term, $\beta \sum_i 1/\phi_i$, to the Flory-Huggins energy function. This modification allows to improve the efficiency of computation, while not changing the rate of morphology evolution or the shape of forming structure. The parameter $\beta$ is set to a small value~$0.001\;RT/V^s$.  }
According to this theory, the energy of the system is given by: 
\begin{equation}
f(\phi_{p},\phi_f,\phi_s)=\displaystyle\frac{RT}{V^{s}}
\displaystyle\left[
\frac{\phi_{p}}{N_{p}}ln(\phi_{p})
+\frac{\phi_{f}}{N_{f}}ln(\phi_{f})
+\frac{\phi_{s}}{N_{s}}ln(\phi_{s})
+\chi_{pf}\phi_{p}\phi_{f}
+\chi_{ps}\phi_{p}\phi_{s}
+\chi_{fs}\phi_{f}\phi_{s}
\displaystyle\right]
\label{eqFloryHuggins}
\end{equation}
where:
$R$ is the gas constant; $T$ is the temperature; $\phi_{i}$ is the volume fraction of component $i$; $N_{i}(=V^i/V^r)$ is the degree of polymerization of component $i$. Here, $V^i$ is the molar volume; $V^r$ is a reference molar volume (e.g. solvent); and $\chi_{ij}$ is the Flory-Huggins binary interaction parameter between component $i$ and $j$. The interaction parameters and degree of polymerization define the shape of the binodal and spinodal lines~(Figure~\ref{fig:PS:scheme}).

One of the advantages of the phase-field method is the ability to introduce additional driving forces to the model in a straightforward way. 
For instance, the effect of the substrate is a surface energy term, $F_s (\phi_p,\phi_f,\textbf{x})$ that is simply added to the  total energy of the system (Eq.~\ref{eq:Energy}). 
A generalized form of the substrate energy is given in Eq.~\ref{eq:fs}. Patterning is introduced through the space dependent functions $p_p(\textbf{x})$ and $p_f(\textbf{x})$ in Eq.~\ref{eq:fs}. 
These functions determine the geometry of patterning.  At a point $\textbf{x}$ on the substrate that is chemically tuned to component $i$, the value  of $p_i=1$, otherwise $p_i=0$.
The chemical specificity of patterning $i$ is reflected in the parameters $\mu^i $ and $h^i$. Parameter $\mu^i$ is a chemical potential favoring component $i$ at the substrate, and $h^p$ expresses the way interactions between the components near the substrate are modified by the presence of a pattern at the substrate~\cite{JonesRichard99,JonesSchwarz89}.
\begin{equation}
F_s(\phi_p,\phi_f,\textbf{x})=-\frac{RT}{V^s}\int_S f_s dS = -\frac{RT}{V^s}\int_S\left[ p_p(\textbf{x})(\mu^{p} \phi_p+h^p \phi_p^2) + p_f(\textbf{x})(\mu^{f}\phi_f+h^{f} \phi_f^2) \right] dS
\label{eq:fs}
\end{equation}

\subsubsection{Governing equations}

Once the energy of the system is specified, the governing equations of the evolution can be formulated. This is usually done by defining the chemical potentials of the system. The chemical potential quantifies how much the energy changes when the configuration changes. 
The chemical potential for polymer and fullerene are $\mu_p=\delta F/\delta \phi_p$ and $\mu_f=\delta F/\delta \phi_f$, respectively.
Next, using Fick's First Law for the flux ($J_p=-M_p\nabla\mu_p$) and the continuity equation ($\partial \phi_p / \partial t  + \nabla J_p = 0$) we get the governing equation for each component.  We consider only two of the three variables as independent (since $\phi_s = 1-\phi_p-\phi_f$). 
The resulting Cahn-Hilliard-Cook equations are given by:
\begin{equation}
\frac{\partial \phi_p}{\partial t}+u\frac{h}{h^{curr}}\frac{\partial \phi_p}{\partial h}
	=\nabla \cdot \left[M_p \nabla \left(
		\frac{\partial f(\phi_p,\phi_f)}{\partial \phi_p} - \epsilon^2 \nabla^2 \phi_p  + (1-H\left(h\right))\frac{\partial f_s(\phi_p,\phi_f)}{\partial \phi_p} 
		\right)\right] + \xi_p
		\label{eq:GovEq1}
\end{equation}

\begin{equation}
\frac{\partial \phi_f}{\partial t}+u\frac{h}{h^{curr}}\frac{\partial \phi_f}{\partial h}
	=\nabla \cdot \left[M_f\nabla \left(
		\frac{\partial f(\phi_p,\phi_f)}{\partial \phi_f} - \epsilon^2 \nabla^2 \phi_f + (1-H\left(h\right))\frac{\partial f_s(\phi_p,\phi_f)}{\partial \phi_f} 
		\right) \right] + \xi_f
		\label{eq:GovEq2}		
\end{equation}
where $M_i=D/(\partial^2f_{ideal}/\partial^2 \phi_i)$ is the mobility of component $i$. 
The diffusivity, $D$, is a linear combination of self-diffusivities of all components and their volume fractions: $D=D_p\phi_p+D_f\phi_f+D_s\phi_s$. 
The energy of ideal solution, $f_{ideal}$, is used to link mobility with diffusivity and to comply with classic Fick's law. 
The ideal solution is one with zero interaction parameters. 

The advection term (second term in LHS of both equations) accounts for the change in height of the film and is scaled according to the height. Scaling term $h/h^{curr}$ is zero at the bottom surface ($h=0$) and one at the top surface ($h=h^{curr}$), where $h^{curr}$ is the total current height of the film.
Substrate effects are included in the third RHS term of both equations. 
This term enters the equation only for the bottom surface (zero height $h=0$) where the Heaviside function, $H(h)=0$.
The last term of RHS (in both equations) is the Langevin force term, $\xi_p$ and $\xi_f$. 
This term mimics the conserved noise due to fluctuations in composition.
We set the noise to be a Gaussian space-time white noise with the following constrains: 
$\langle\xi_i(t_1,\textbf{x}_1)\rangle=0$ and 
$\langle\xi_i(t_1,\textbf{x}_1) \xi_i(t_2,\textbf{x}_2)\rangle=2M_iRT/V^s\delta(t_1-t_2)\nabla^2\delta(\textbf{x}_1-\textbf{x}_2)$, where $\delta$ is the Kronecker delta.
Variance is determined by the fluctuation-dissipation theorem (FDT)~\cite{HohenbergHalperin1977}. 
Cahn-Hilliard equation with noise considered is called the Cahn-Hilliard-Cook equation~\cite{Cook70}.

This formulation allows for a natural extension to include crystallization. To do this, an additional phase variable must be considered, the energy functional  expanded accordingly and the governing equation formulated. 

\subsection{Modeling solvent evaporation}

From a mathematical perspective, an evaporation process is classified as a moving boundary problem, since the height of the thin film changes over time ($h=h(t)$). 
Following~\cite{KSM09}, we explicitly trace the surface evolution.
We assume that solvent is the only component that evaporates from the top surface. 
Solvent lost from the top layer results in height decrease. 
The rate at which the height decreases, $u=\partial h/\partial t$, is proportional to the flux of the solvent out of system, $J_s^a$:
\begin{equation}
\frac{\partial h}{\partial t}=-V^s J_s^a
\label{eq:u}
\end{equation}
where $J_s^a$ is the normal component of molar flux (of solvent into the air). 
This equation constitutes the mass balance for moving film-air interface: $u(\phi_i^a-\phi_i^s)=V^s(J_i^a-J_i^s)$ with two assumptions: only solvent evaporates ($J_p^a=J_f^a=0$) and solvent content in the air is very low ($\phi_s^a\ll1$). The superscript $a$ and $s$ denote air and film, respectively. The molar flux of solvent can be further linked with evaporation rate of the solvent, $k_e$, and the average content of the solvent at the top layer, $\phi_s^{top}$, using equation:
\begin{equation}
V^s J_s^a=k_e\bar{\phi}_s^{top}.
\end{equation}
In this work, we assume that the solvent evaporates uniformly from the top surface and film height decreases homogenously. This assumption simplifies analysis and quantification of the competing effects of evaporation and substrate shown in the results section. We investigate the effect of inhomogeneous evaporation on the film surface evolution in a forthcoming publication.

The evaporation rate, $k_e$, depends on various parameters including the solvent partial pressure, solvent vapor pressure, temperature, air flow and is specific to the fabrication process. 
For spin-coating, the evaporation rate depends on the angular velocity $k_{e}=k_{e}(\omega^{1/2})$~\cite{Meyerhofer78}, while for solvent annealing or drop-casting, the evaporation rate can be assumed constant ($k_{e}=const$). 
We apply boundary conditions at the top surface to satisfy the balance of two other components within the film. Removal of the solvent from the top layer, results in enrichment of the volume fraction of polymer and fullerene. Therefore, to account for this enrichment we apply Neumann boundary conditions at top surface for two other components: $J_{i\neq s}=-J_{s}^{a}\phi_{i\neq s}$.
\footnote{Applied boundary conditions guarantee that content of polymer and fullerene is conserved, that was confirmed by thorough tests performed.}

\subsection{Challenges}
Although the phase field method is a well used technique for simulating the morphological evolution of a wide variety of materials and  processes~\cite{Cogswell2010,Emmerich03,SKP07}, employing it to predict morphology evolution during fabrication of active layer for organic solar cells requires resolving several challenges. We detail these challenges below.

\subsubsection{Physical Modeling Challenges}
Morphology evolution during fabrication of organic solar cells is an {\it intrinsically multiscale process both in time and space} which makes this process difficult to solve accurately and efficiently using reasonable computational resources. The governing equations (Eqns.~\ref{eq:GovEq1}, \ref{eq:GovEq2}) are forth-order nonlinear partial differential equations, which are difficult to solve numerically. 
The complexity of these equations is related to two competitive subprocesses: {\it phase separation and coarsening}. 
These processes occur at two widely different temporal and spatial scales, all of which must be resolved properly.
Initially, very fast phase separation is the dominant process and followed by slow coarsening. 
Phase separation is a fast process and results in thin layer creation. 
In contrast, coarsening is slow process consisting of rare events and involving merging of bulky regions.

Another numerical challenge is related to the {\it evaporation process}. 
In case of the solvent-based thin-film deposition, the volume fraction of solvent changes from 99\% to almost zero. This poses a severe problem for numerical techniques that have to reliably model the huge change in domain size in 3D. 

\subsubsection{Computational Challenges} 
A key objective of the formulation is the ability to predict morphology evolution at \emph{the device scale}. The inherent complexity of the process makes this objective demanding. This is because we are interested in resolving nano scale morphological evolution while investigating device-scale domains.  
From a computational perspective this involves solving differential equations with a very large number of degrees of freedom ($\gg10^{7}$), which cannot be solved using current serial processing machines.  This require developing modules heavily based on parallel processing, including applying domain decomposition strategies~\cite{SKK00,SKK02}.

\subsubsection{Materials Science Challenges} 
The phase field method is a generic technique, and can be applied to almost any type of system undergoing morphology evolution. 
Thus, in order to provide quantitative prediction, it is necessary to determine material-specific set of parameters, both thermodynamic and kinetic. 
These parameters very often show compositional, directional or temperature dependence, which pose additional difficulties. In this regard, significant work has been done to determine several parameters for materials and systems utilized for photovoltaic applications using both molecular dynamics~\cite{HFD10,HMF10} and experimental techniques~\cite{JRB07,NBB07,SHSK10}.

\section{Numerical model}

\subsection{Strong form of the split Cahn-Hilliard equations with boundary conditions}
	The strong form is formulated as follows: find $\phi_p, \phi_f:\Omega \times [0,T]\to \mathbb{R}$ and $\mu_p, \mu_f:\Omega \times [0,T]\to \mathbb{R}$ ($\mu$ is an auxiliary variable) such that:

	\begin{eqnarray}
	\frac{\partial \phi_p}{\partial t}+u\frac{h}{h^{curr}}\frac{\partial \phi_p}{\partial h} = \nabla \cdot \left(M_p\nabla \mu_p\right) +  \xi_p    	& in & \;\Omega   \times [0,T], \label{eq:sfCHpa}\\
	\frac{\partial \phi_f}{\partial t} +u\frac{h}{h^{curr}}\frac{\partial \phi_f}{\partial h} =\nabla \cdot \left(M_f\nabla \mu_f\right) +  \xi_f       & in & \;\Omega   \times [0,T], \label{eq:sfCHfa}\\
	\mu_p = \displaystyle \frac{\partial f}{\partial \phi_p} - \epsilon^2 \nabla^2 \phi_p  + \left(1-H(h)\right)\frac{\partial f_s}{\partial \phi_p}  	& in & \;\Omega   \times [0,T], \label{eq:sfCHpb}\\
	\mu_f= \displaystyle \frac{\partial f}{\partial \phi_f} - \epsilon^2 \nabla^2 \phi_f  + \left(1-H(h)\right)\frac{\partial f_s}{\partial \phi_f}  	& in & \;\Omega   \times [0,T], \label{eq:sfCHf}\\
	\nabla (M_p\mu_p) \cdot \textbf{n} =  h_{\mu_p}				       																					& on & \;\Gamma_t \times [0,T], \label{eq:sfCHpe}\\
	\nabla (M_f\mu_f) \cdot \textbf{n} =  h_{\mu_f}				        																				& on & \;\Gamma_t \times [0,T], \label{eq:sfCHfe}\\
	\phi_p(\textbf{x},0)=\phi_p^0(\textbf{x})					        																				& in & \;\Omega, \\
	\phi_f(\textbf{x},0)=\phi_f^0(\textbf{x})					        																				& in & \;\Omega.
	\end{eqnarray}
We apply Neumann boundary conditions on the top surface, $\Gamma_t$, to account for polymer and fullerene fraction increase at the top layer due to solvent evaporation. Flux of polymer $h_{\mu_p}$ equals to $-J_{s}^a\phi_{p}$, while flux of the fullerene $h_{\mu_f}$ equals to $-J_{s}^a\phi_{f}$. On other boundaries, we apply zero flux conditions for both variables: volume fraction, $\phi_i$,  and chemical potential, $\mu_i$.

    \subsection{Domain rescaling due to evaporation}
	To account for domain change due to solvent removal from the top layer, we introduce a coordinate transformation~\cite{ShojaieGreenberg1994,SaylorWarren2011}:
	\begin{equation}
	\vartheta=\frac{h}{h^{curr}(t)}
	\label{eq:eta}
	\end{equation}
	where $h$ is the height coordinate scaled according to current total height of the film, $h^{curr}$. 
	This coordinate transformation permits recasting the model equations into a system of equations with fixed boundaries which is more convenient for numerical solution. \emph{In this way, there is no need for remeshing.} 

Since we assume homogeneous evaporation, film has an uniform height $h^{curr}(t)$, at any point in time, $t$. Recasting the model equation using the coordinate transformation~\ref{eq:eta} involves defining the gradient operator in the new coordinate system:
	\begin{equation}
	\tilde{\nabla}=\hat{e}_1\frac{1}{h^{curr}} \frac{\partial}{\partial \vartheta}+\hat{e}_i\frac{\partial}{\partial x_i}
	\end{equation}
	where the first component is the direction along the height of the film. Strong forms of the recast equations are given  by:
		\begin{eqnarray}
	\frac{\partial \phi_p}{\partial t}+u\frac{\vartheta}{h^{curr}}\frac{\partial \phi_p}{\partial \vartheta}&=&\tilde{\nabla}\cdot\left(M_p\tilde{\nabla}\mu_p\right)+\xi_p\qquad in\;\Omega   \times [0,T], \label{eq:t:sfCHpa}\\
	\frac{\partial \phi_f}{\partial t}+u\frac{\vartheta}{h^{curr}}\frac{\partial \phi_f}{\partial \vartheta}&=&\tilde{\nabla}\cdot\left(M_f\tilde{\nabla}\mu_f\right)+\xi_f\qquad in\;\Omega   \times [0,T], \label{eq:t:sfCHfa}\\
	\mu_p&=&\displaystyle \frac{\partial f}{\partial \phi_p} - \epsilon^2 \tilde{\nabla}^2 \phi_p  + \left(1-H(h)\right)\frac{\partial f_s}{\partial \phi_p} \qquad in\;\Omega   \times [0,T], \label{eq:t:sfCHpb}\\
	\mu_f&=&\displaystyle \frac{\partial f}{\partial \phi_f} - \epsilon^2 \tilde{\nabla}^2 \phi_f  + \left(1-H(h)\right)\frac{\partial f_s}{\partial \phi_f}  	\qquad in\;\Omega   \times [0,T], \label{eq:t:sfCHf}\\
	\tilde{\nabla} (M_p\mu_p) \cdot \textbf{n}&=& h_{\mu_p}				        	\qquad on\;\Gamma_t \times [0,T], \label{eq:t:sfCHpe}\\
	\tilde{\nabla} (M_f\mu_f) \cdot \textbf{n}&=& h_{\mu_f}				        	\qquad on\;\Gamma_t \times [0,T], \label{eq:t:sfCHfe}
	\end{eqnarray}
The advection term (second LHS term of Eqns~\ref{eq:t:sfCHpa} and~\ref{eq:t:sfCHfa}) is non zero only along the height direction. Velocity corresponding to this term, $u$, is proportional to the molar flux of the solvent out of system: $u= -V^s J_s^a = k_e \bar{\phi}_s^{top}$ ( Eq.~\ref{eq:u}).

    \subsection{Spatial discretization of governing equation}
	We use the finite element method to solve the governing equations (Eqns~\ref{eq:t:sfCHpa} - \ref{eq:t:sfCHfe}). We solve equations in the split form, to avoid constraints related to the continuity of the basis functions when using the finite element method with a primal variational formulation~\cite{EFM89}. More precisely, the standard $\mathcal C^0$-continuous finite element formulation is not suitable for forth-order operators, and consequently basis functions which are piecewise smooth and $\mathcal C^1$-continuous should be utilized~\cite{WKG06,GCB08}. However, there are only a limited number of finite elements that fulfill the above continuity condition, especially in two and three dimensions.	

	The weak form of the split Cahn-Hilliard equation is given by:
\begin{equation}
\left(w,\phi_{p,t}\right)_{\Omega}+\left(w, u_i\vartheta/h^{curr} \phi_{p,i}\right)_{\Omega}+a\left(w,M_p\mu_p\right)_{\Omega}+a\left(w,\rho_{p}\right)_{\Omega}=\left(w,h_{\mu_p}\right)_{\Gamma_t}
\label{eq:wfCHap}
\end{equation}
\begin{equation}
\left(w,\phi_{f,t}\right)_{\Omega}+\left(w, u_i\vartheta/h^{curr} \phi_{f,i}\right)_{\Omega}+a\left(w,M_f\mu_f\right)_{\Omega}+a\left(w,\rho_{f}\right)_{\Omega}=\left(w,h_{\mu_f}\right)_{\Gamma_t}
\label{eq:wfCHaf}
\end{equation}
\begin{equation} 
\displaystyle\left(w, \frac{\partial f}{\partial \phi_p}\right)_{\Omega}-\left(w,\mu_p\right)_{\Omega}+a\left(w,\epsilon^2 \phi_p\right)_{\Omega} + \left(w, \frac{\partial f_s}{\partial \phi_p}(\phi_i)\right)_{\Gamma_b} =0
\label{eq:wfCHbp}
\end{equation}
\begin{equation} 
\displaystyle\left(w, \frac{\partial f}{\partial \phi_f}\right)_{\Omega}-\left(w,\mu_f\right)_{\Omega}+a\left(w,\epsilon^2 \phi_f\right)_{\Omega} + \left(w, \frac{\partial f_s}{\partial \phi_f}(\phi_i)\right)_{\Gamma_b} = 0
\label{eq:wfCHbf}
\end{equation}
where $w\in \mathcal H^1(\Omega)$ are weighting functions, $(\cdot,\cdot)$ is the $L_2$ inner product on $\Omega$, $a(\cdot,\cdot)$ is the energy inner product on $\Omega$, $h_{\phi_i}$ and $h_{\mu_i}$ define natural boundary conditions. 
Second term in LHS of Eqns~\ref{eq:wfCHap} and~\ref{eq:wfCHaf} accounts for domain size change due to evaporation. 
Fourth term in LHS of Eqns~\ref{eq:wfCHap} and~\ref{eq:wfCHaf} accounts for the conserved noise, where $\rho_i$ is the vector of the stochastic flux terms.
Fourth term in LHS of Eqns~\ref{eq:wfCHbp} and~\ref{eq:wfCHbf} accounts for substrate effect and are included only for surface elements belonging to bottom boundary $\Gamma_b$. 
We note that substrate term is not a typical boundary condition but an additional term resulting from the additional energy in the system.

We use the Galerkin approximation to solve the two split Cahn-Hilliard equations. We define $\phi_p^h \in \mathcal S_p^h$ and $\phi_f^h \in \mathcal S_p^h$ to be the finite dimensional approximation of polymer and fullerene volume fraction fields, $\mu_p^h \in \mathcal M_p^h$ $\mu_f^h \in \mathcal M_f^h$ to be the finite dimensional approximation of polymer and fullerene chemical potential fields and $w^h \in \mathcal V^h$ to be the finite dimensional weighting function. The approximate solutions are computed on the following function spaces:

\begin{equation}
\mathcal S_p^h = \{\phi_p^h | \phi_p^h \in \mathcal H^1(\Omega),\; \phi_p^h \in P^k(\Omega^e)\; \forall e\} 
\end{equation} 
\begin{equation}
\mathcal M_p^h = \{\mu_p^h | \mu_p^h \in \mathcal H^1(\Omega),\; \mu_p^h \in P^k(\Omega^e)\; \forall e\} 
\end{equation} 
\begin{equation}
\mathcal S_f^h = \{\phi_f^h | \phi_f^h \in \mathcal H^1(\Omega),\; \phi_f^h \in P^k(\Omega^e)\; \forall e\} 
\end{equation} 
\begin{equation}
\mathcal M_f^h = \{\mu_f^h | \mu_f^h \in \mathcal H^1(\Omega),\; \mu_f^h \in P^k(\Omega^e)\; \forall e\} 
\end{equation} 
\begin{equation}
\mathcal V^h = \{w^h | w^h \in \mathcal H^1(\Omega),\; w^h \in P^k(\Omega^e)\; \forall e\} 
\end{equation} 
with $P^k(\Omega^e)$ being the space of the standard polynomial finite element shape functions on element $\Omega^e$, where $k$ is the polynomial order. 
We additionally introduce the SUPG stabilization term for the advection terms in Eqs~\ref{eq:wfCHap} and~\ref{eq:wfCHaf}~\cite{Hughes1982}.

We discretize the conserved noise by generating Gaussian- distributed random numbers for each component of flux $\hat{\rho}$ that satisfy $\langle\hat{\rho_i}(t_1,x_1)\rangle=0$ and 
$\langle\hat{\rho_i}\hat{\rho_{i'}}\rangle=1$. 
We generate $n_{sd}$-dimensional vector ($n_{sd} = 2, 3$ depending on 2D or 3D problem) for each component (polymer and fullerene) in each node at each time step, $\rho_{pi}=\varrho_p\hat{\rho_i}$. 
Parameters $\varrho_p=\sqrt{2M_p RT/(V^s \Delta t)}$ and $\varrho=\sqrt{2M_f RT/(V^s \Delta t)}$ (where $\Delta t$ is size of the time step) account for fluctuation-dissipation theorem requirement.
For more details see~\cite{ShenWang2007,KarmaRappel1996}.

We use linear basis functions for variables $\phi_p$, $\phi_f$, $\mu_p$, $\mu_f$, $\hat{\rho_p}$ and $\hat{\rho_f}$.  We also tested quadratic and cubic basis functions but have not noticed any significant improvements. This choice is based on an extensive analysis of the Cahn-Hilliard equation in~\cite{WG11a}.

    \subsection{Temporal discretization of governing equation}

	In our approach, we take advantage of implicit time schemes due to their unconditional stability. Explicit methods are often intractable due to severe restrictions on size of time step ($\sim\Delta x^4$) arising from the stiffness of the equation. 
This makes them computationally prohibitive even for simple problems. 
Consequently, implicit methods arise as a natural alternative. Since they are unconditionally stable they allow for much larger time step. 
However, because of nonlinear nature of the Cahn-Hilliard equation, implicit schemes require nonlinear solvers. 
The popular implicit time schemes in this context are Euler Backward and Crank-Nicholson methods~\cite{Khiari2007,RFKS10,WKG06}. 

Because of the multiple temporal scales that occur during phase separation, we tested adaptive various time stepping strategies~\cite{WG11a}. We noticed significant improvement in terms of number of times steps required to reach steady state as well as in total run times~\cite{WG11a}. 
In these time stepping strategies, step size is adjusted on the basis of the error between solutions of different order. However, whenever noise is considered, the error computed using such strategies is highly affected by the noise and cannot be used in standard time stepping strategies. Therefore, we use a Euler Backward scheme with a heuristic strategy to adjust the time step as used in~\cite{RFKS10} (see~\ref{sec:app}). 

\section{Technical details}
To solve the nonlinear system of two split Cahn-Hilliard equations, the formulation is linearized consistently and a Newton--Raphson scheme is used. 
To solve large problems with several millions of degrees of freedom, we use a domain-decomposition based mesh-partitioner to divide the mesh and distribute it across computational nodes.  
In our framework we use the ParMETIS partitioner~\cite{SKK00,SKK02}.  
Additionally, to enable parallel solution of the algebraic systems, we use the PETSc solver library~\cite{petsc-web-page,petsc-user-ref}.  
All results reported are obtained using the Generalized Minimal Residual Method. 

\section{Results}
\label{ch:RD}
In this section, we showcase the formulation by investigating the effect of evaporation rate,  blend ratio,  degree of polymerization as well as effect of choice of solvent on morphology evolution. Finally, we investigate the possibility of additional control of morphology through substrate patterning. 

We investigate a representative OSC system. 
Such a system consists of polymer, fullerene and solvent. 
In the default configuration, we consider degree of polymerization of solvent $N_p=5$, fullerene $N_f=5$ and solvent $N_s=1$.
Degree of polymerization is computed on the basis of molar volumes of the components. 
We assume $V^s=100\;cm^3/mol$, $V^f=500\;cm^3/mol$, $V^p=500\;cm^3/mol$.
Interaction parameter between polymer and fullerene reflects the low solubility of two components and is set as$\chi_{sf}=1.0$.
Interaction parameters between solvent and fullerene or polymer are assumed to be much lower: $\chi_{fs}=0.3, \chi_{ps}=0.3$.
In subsection~\ref{subsec:N}, we investigate the effect of varying degree of polymerization to $N_p=20$ and $N_p=100$.
In subsection~\ref{subsec:solvent}, we investigate the effect of solvent and various interaction parameters.
All these values are representative for OSC and correspond well with experimentally determined values~\cite{NBB07}.
We take the solvent self-diffusion coefficient as $D_s=10^{-9}m^2/s$, which are typical for solvents (chlorobenzene, chloroform and xylene) used in OSC~\cite{GerekElliot2010}. 
Self-diffusion of polymer and fullerene is much lower. We assume that $D_p=D_f=10^{-3} D_s$.

We present one, two and three dimensional results. 
For each simulation we generate one mesh in the reference coordinate system. 
Mesh density is adjusted to the estimated width of the interface between polymer and fullerene. We use the interface width as a metric to determine mesh density to accurately capture the dynamics of the interface evolution.  Our detailed analysis presented in~\cite{WG11a} showed that at least four elements per interface are required to capture the physics of phase separation and coarsening accurately. 

In~\cite{WG11a}, we showed that the analytical estimation of interface width is fairly accurate for two and three dimensional cases, even though it was derived for the one dimensional case in~\cite{CH58a}. The interface width is defined as a distance required to span by the concentration profile across the accessible range of the composition: $\delta=\Delta \phi_e/(d\phi/dx)|_{\phi_c}=\Delta \phi_e \sqrt{\epsilon^2/\Delta f_{max}}$. For more details see~\cite{CH58a,WG11a}.
The profile of concentration, and subsequently width of the interface, depends on the interfacial parameter $\epsilon$, interaction parameters, degree of polymerization. 
In ternary system range of the composition changes with time at the intermediate stages. 
Consequently, the interface width also changes with evaporation.
To estimate interface width, we consider the case when the interface width is the smallest. This corresponds to the fully evaporated, binary fullerene-polymer system. For the system modeled in this paper, the interface width between polymer and fullerene varies from $12\;nm$ (for $N_p=5$) to $6\;nm$ (for $N_p=100$). 

The interfacial parameter $\epsilon$ is also estimated for a binary system consisting of polymer and fullerene.  Following the analysis in~\cite{CH58a,SKP07}, one can link $\epsilon$ with the interfacial energy, $\gamma_{pf}$,  between polymer-rich and fullerene-rich phases. 
We assume $\gamma_{pf}=33\;mJ/m^2$, which is comparable with interfacial energies for organic compounds.
For more details regarding detailed analysis and computations, we refer the reader to~\cite{SKP07}.
Computed value of $\epsilon^2$ is $3.57\cdot10^{-10}J/m$ (for $N_p=5$), $2.62\cdot10^{-10}J/m$ ($N_p=20$) and $\epsilon^2=2.05\cdot10^{-10}J/m$ ($N_p=100$). 

The initial solvent fraction is chosen to guarantee that the ternary solution is homogeneous and there is no phase separation at time $t=0$.
In 3D simulations, we start with a  solvent fraction $\phi_s=0.66$. 
In 2D simulations, we start with a  solvent fraction $0.75$.
In 1D simulations, we start with a solvent fraction $0.6$.
The higher initial volume fraction of solvent in 2D simulations is required to cover the wide range of degree of polymerization and interaction parameters investigated. 
Simulation is stopped when the fraction of solvent within the film is $0.05$. 
At this time, the diffusion coefficient reduces significantly, and the morphology is frozen.

In two dimensional simulation, we generate meshes that consist of $250\times100$ linear elements. The computational domain is rectangular with height $L_y=1$, and width $L_x=2.5$. 
Height changes from $Ly=1$ to $0.26$, when all the solvent evaporates.
In three dimensional simulations, we generated meshes that consist of $230\times230\times70$ linear elements to discretize a computational domain that is a rectangular prism of length $L_x=3.3$, breadth $L_y=3.3$ and height $L_z=1.0$. Height changes from $L_z=1$ to $0.34$, as the solvent evaporates. In one dimensional simulations, we generate meshes that consist of $100$ linear elements to discretize a height $L_x=1$. Height changes from $1.0$ to $0.42$. 

The large number of degree of freedom ($\sim$15 million for 3D simulation) require using parallel solvers and domain decomposition. The average run time for a three dimensional case is around 50h solved using 256CPUs. The average run time for two dimensional case is around 1.5h using 8CPUs. Detailed scalability analysis of the solver has been reported elsewhere~\cite{WG11a}.

\subsection{Influence of evaporation rate}
Evaporation of the solvent from the top surface is one of the key phenomena that induces phase separation. 
The way morphology evolves is an interplay between two competing processes: (i)~evaporation of the solvent from the top surface, and (ii)~the diffusion of the solvent within the film. 
This interplay can be expressed as a mass Biot number. 
Biot number is defined in Eq.~\ref{eq:Bi} and expresses the ratio between external mass transport by evaporation of the solvent from the top surface and internal mass transfer to the top surface by diffusion. 
\begin{equation}
	Bi=\frac{k_e}{D/L} 
\label{eq:Bi}
\end{equation}
where $L$ is the characteristic length - height of the film, $D$ is the diffusion coefficient of the thin film and $k_e$  is the mass transfer coefficient - evaporation rate.
We compute the Biot number for initial height and solvent diffusion coefficient. In such case, the solution consists mostly of solvent, and the diffusion is dominated by it.

\begin{figure}
\centering
\parbox{0.5\textwidth}{\includegraphics[width=0.5\textwidth]{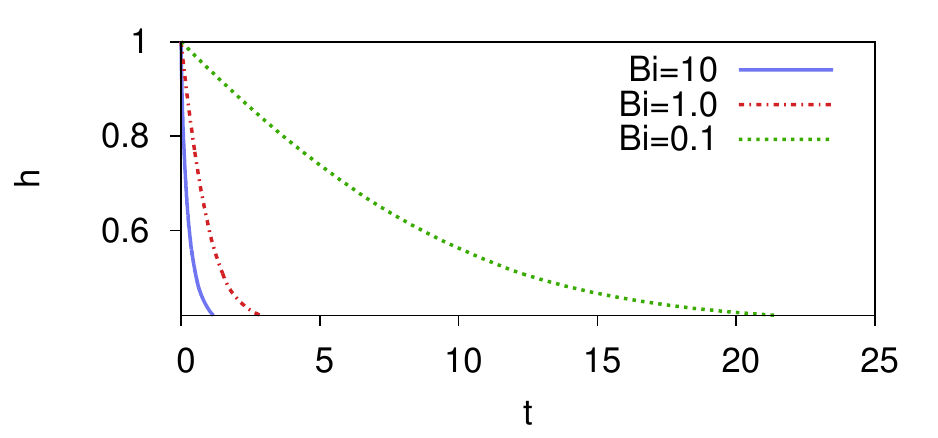}} \\
\parbox{0.32\textwidth}{\centering $Bi=0.1$}
\parbox{0.32\textwidth}{\centering $Bi=1.0$}
\parbox{0.32\textwidth}{\centering $Bi=10$}\\
\parbox{0.32\textwidth}{\includegraphics[width=0.3\textwidth]{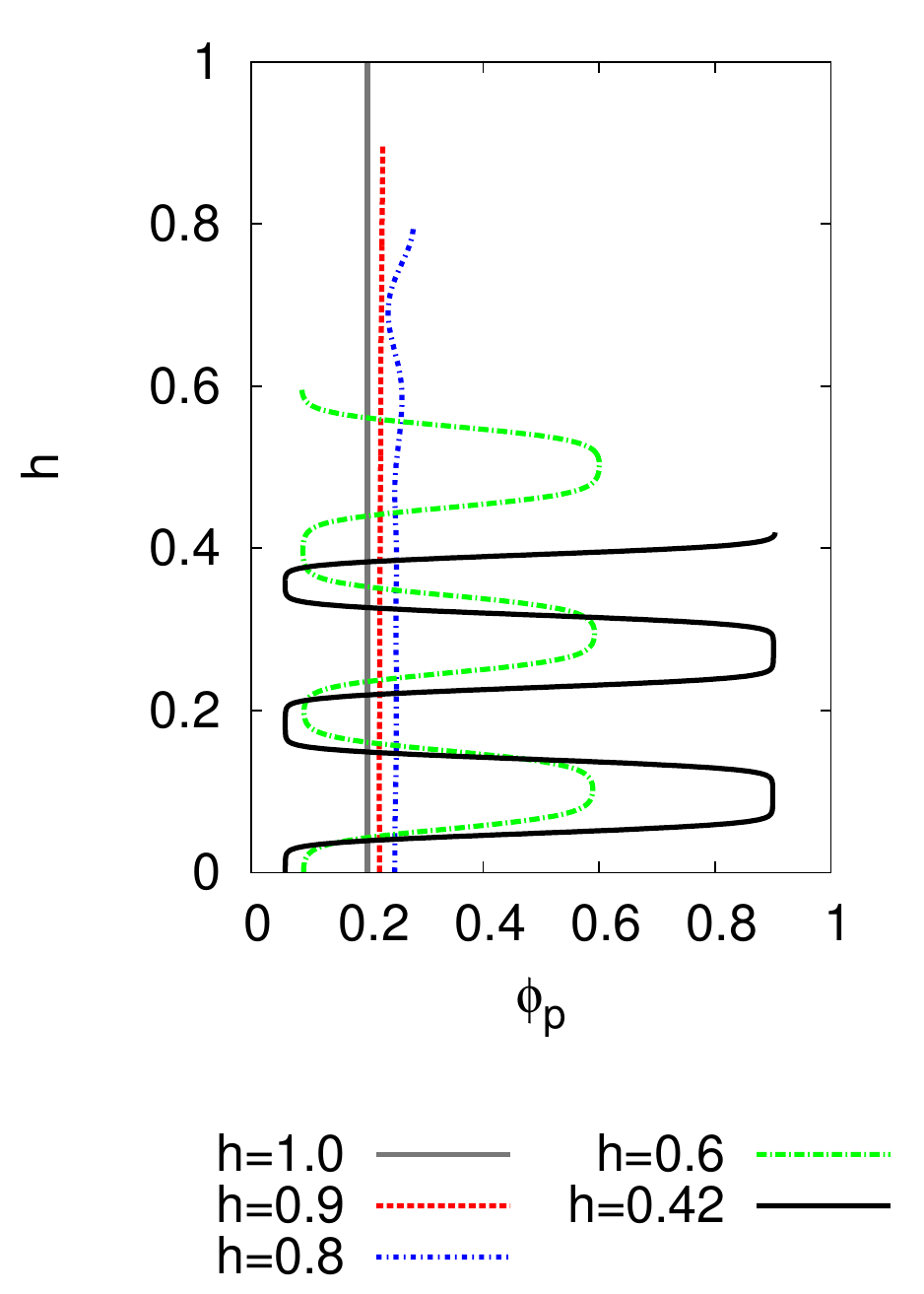}}  
\parbox{0.32\textwidth}{\includegraphics[width=0.3\textwidth]{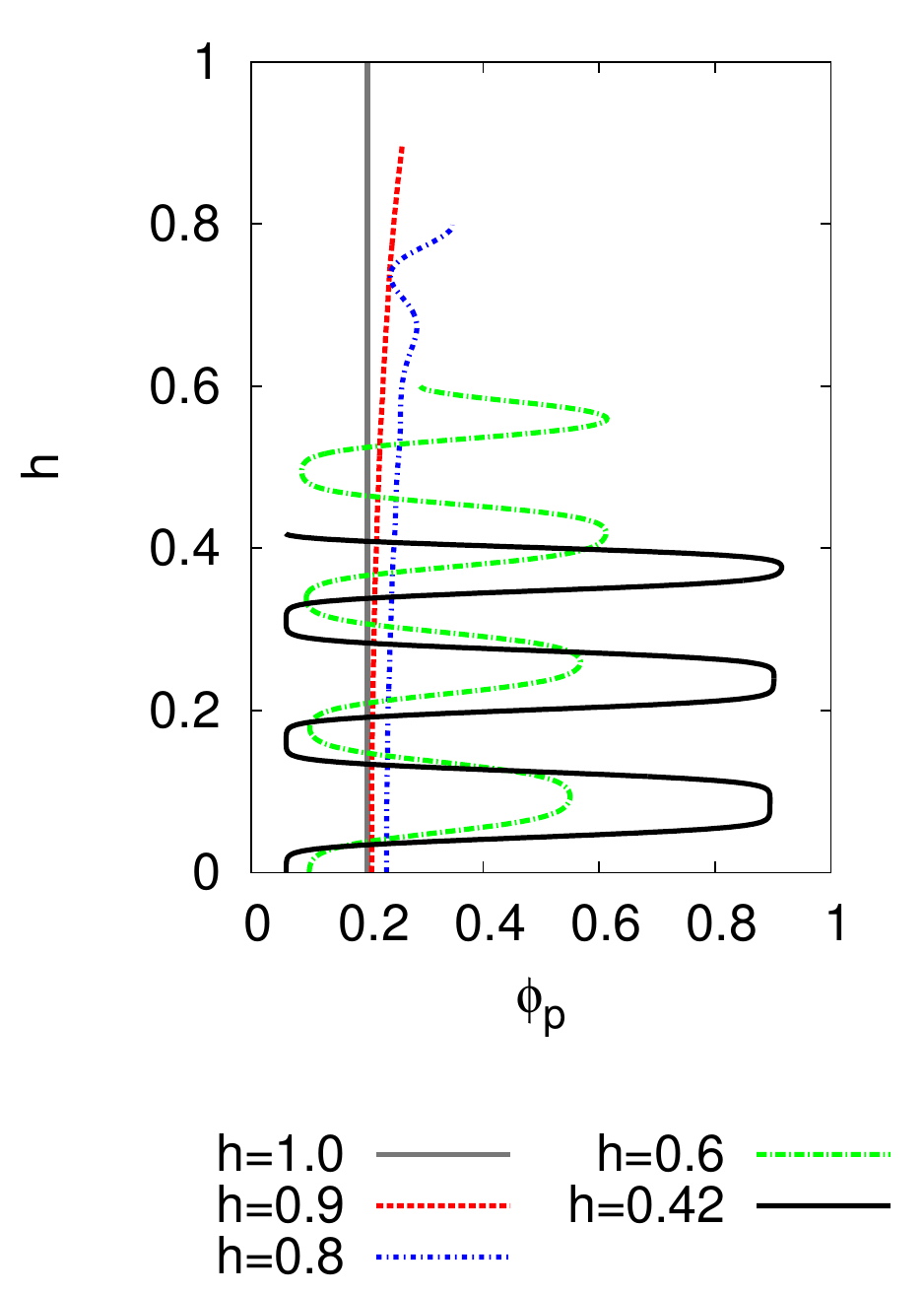}}  
\parbox{0.32\textwidth}{\includegraphics[width=0.3\textwidth]{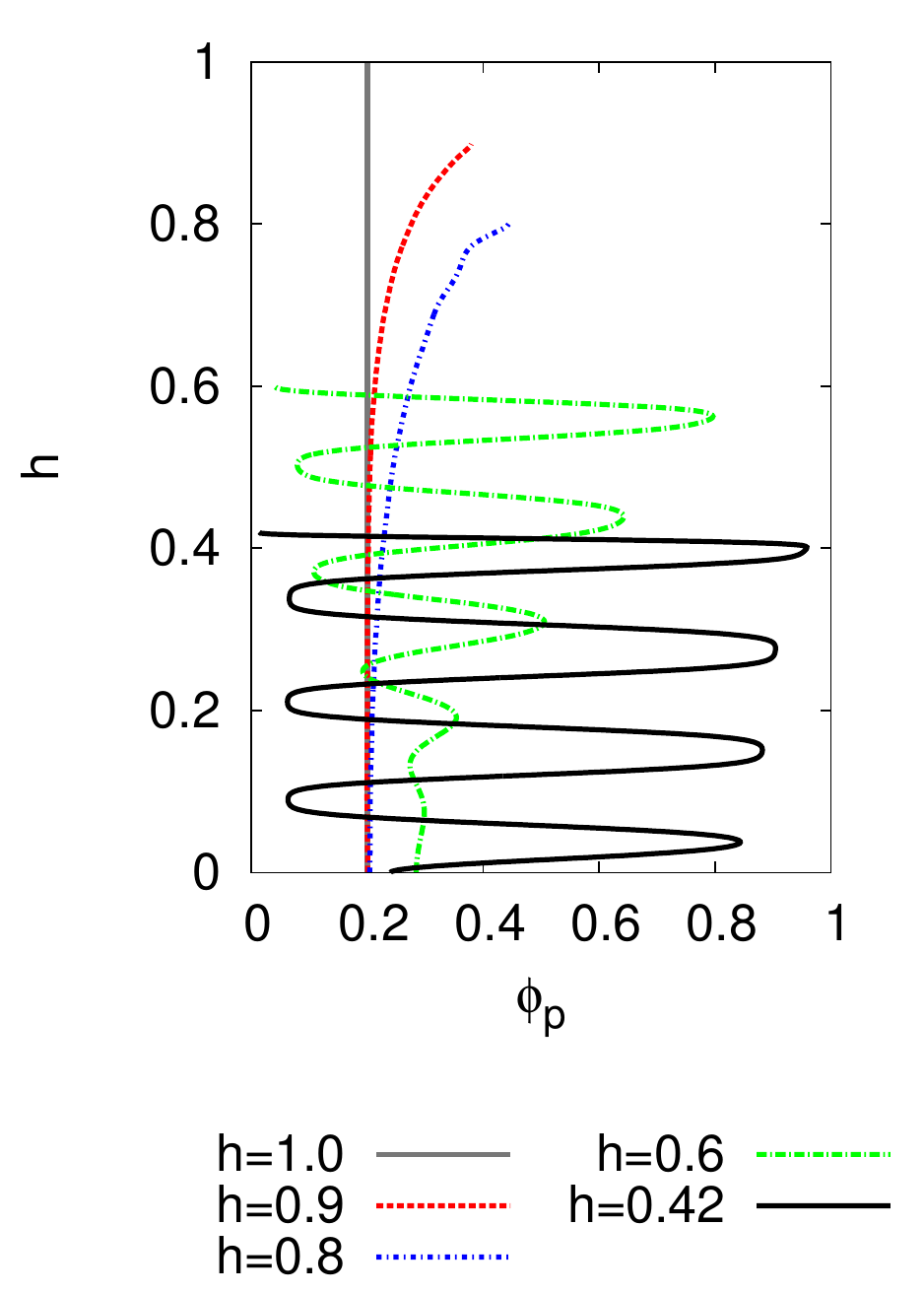}}
\caption{(Top) Change in film height in time for three Biot numbers: $Bi=0.1$, $1$ and $10$. (Bottom) Volume fraction profiles of polymer along film height for three different Biot numbers.}
\label{fig:res:Bi}
\end{figure}

To better understand the interplay between evaporation and diffusion, we first perform experiments in 1D.
We consider the default case of system variables ( $N_p=N_f=5$, $N_s=1$, $\chi_{pf}=1$, $\chi_{ps}=\chi_{fs}=0.3$).  In Figure~\ref{fig:res:Bi}, we plot change of height for three Biot numbers as a function of time.  In the same figure we also show the 1D volume fraction profiles of one component, polymer, as function of height. Initially, polymer is distributed uniformly along the height and its volume fraction is 0.2. Similarly, the initial volume fraction of fullerene is 0.2, since the blend ratio between polymer and fullerene 1:1. For the symmetric case, the fullerene profiles mirror images of polymer profiles (and hence are not plotted).  
We investigate the cases with three Biot number: equal to one, much larger than one ($Bi=10$), and much lower than one ($Bi=0.1$). For this specific system, Biot number $Bi=1$ correspond to initial height $1000\;nm$ and evaporation rate $k_e=0.001\;m/s$.

When Biot number is much larger than one, evaporation is the dominant process. 
This is reflected in shorter total time of evaporation compared to other cases (see Figure~\ref{fig:res:Bi} top). Consequently, the solvent removed from the top layer cannot be balanced by mass diffusion within the film. Thus, a boundary layer lean in solvent and rich in other components is created close to the top (Figure~\ref{fig:res:Bi} (right)).
Enrichment of polymer and fullerene results in the initiation of phase separation. 
In this way, boundary layer becomes the region of the thin film when solution is unstable even to small fluctuations, which leads to phase separation. Phase separation subsequently propagates into the depth of the film. This is clearly seen in Figure~\ref{fig:res:Bi} (right). 
Close to the top layer, a blocking layer rich in polymer is created. Because of low diffusion coefficient of polymer, the top layer blocks solvent evaporation from the top.

When Biot number is much smaller than one, diffusion is the dominant process. Correspondingly, the total time of solvent removal is much longer. The system has more time to balance solvent lost from the top layer.  Diffusion is not suppressed by fast solvent removal (as for high Biot numbers) and consequently no top boundary layer rich in polymer is created. There is no significant gradient of the solvent in the composition within the film.  Consequently, phase separation in initiated homogeneously along the height. 

Biot number provides important insight into interplay between evaporation and diffusion. 
It can be used to link two types of morphology evolution: homogeneous across the film or initiated close to the top surface.
For symmetric systems ($N_p=N_f$) and low Biot number, solvent content within the film is homogeneous. For such a scenario, an assumption of homogeneous decrease of solvent volume fraction {\it within the entire volume is valid}. Such assumption was made in a recent work~\cite{ShangKazmer2011}. However, if one of the non-solvent components has larger molar volume, and consequently large $N_i$ -- which is the case in OSC -- this assumption is invalid and evaporation must be included in the model explicitly. 

\begin{figure}
\begin{flushright}
\includegraphics[width=0.15\textwidth]{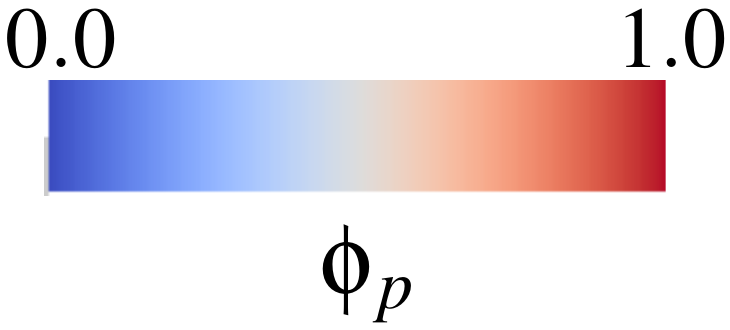}\\
\end{flushright}
\parbox{0.05\textwidth}{
\setlength{\unitlength}{2mm}
\begin{picture}(10,80)
\put(0,50){$t$}
\thicklines
\put(3,65){\vector(0,-1){50}}
\end{picture}
}
\parbox{0.31\textwidth}{\centering $Bi=0.03$ \\
\includegraphics[width=0.3\textwidth]{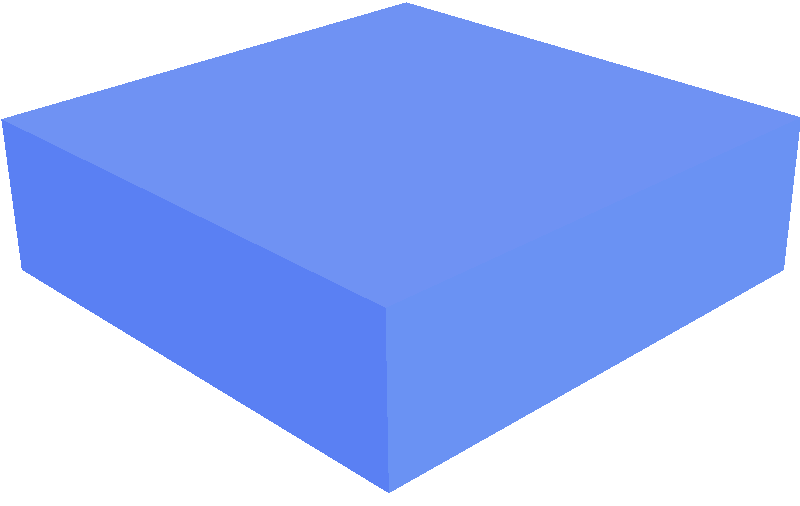}\\
\includegraphics[width=0.3\textwidth]{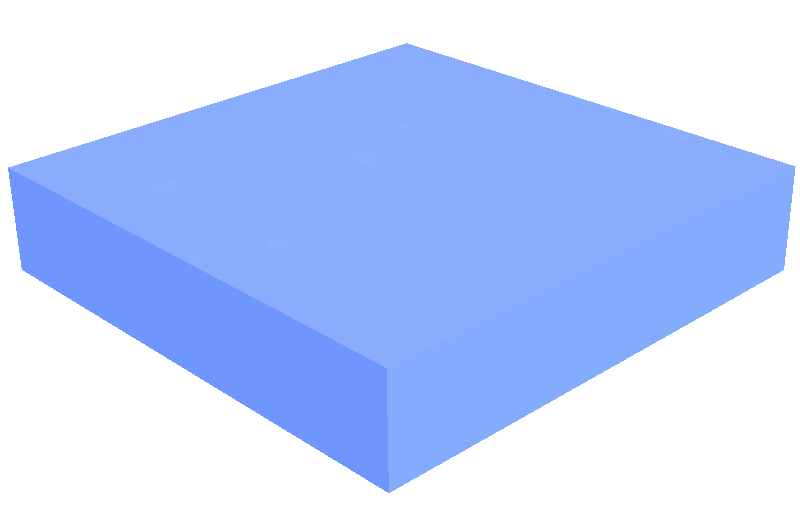}\\
\includegraphics[width=0.3\textwidth]{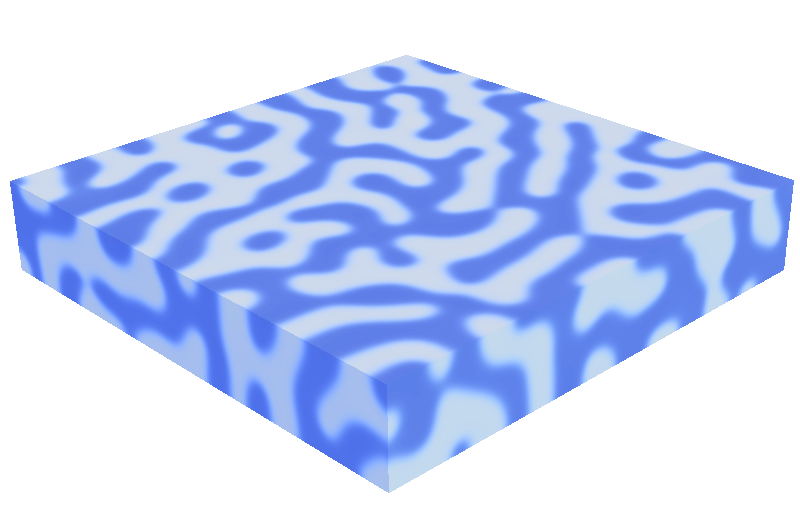}\\
\includegraphics[width=0.3\textwidth]{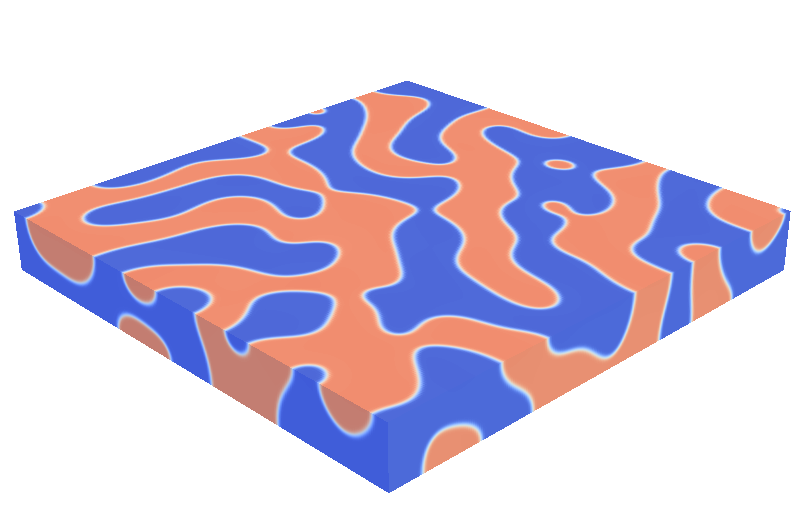}\\
\includegraphics[width=0.3\textwidth]{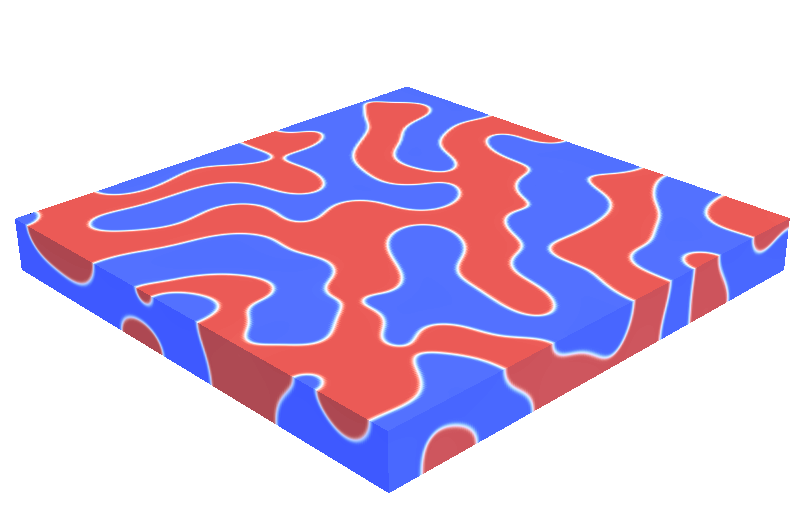}
}
\parbox{0.31\textwidth}{\centering $Bi=0.3$\\
\includegraphics[width=0.3\textwidth]{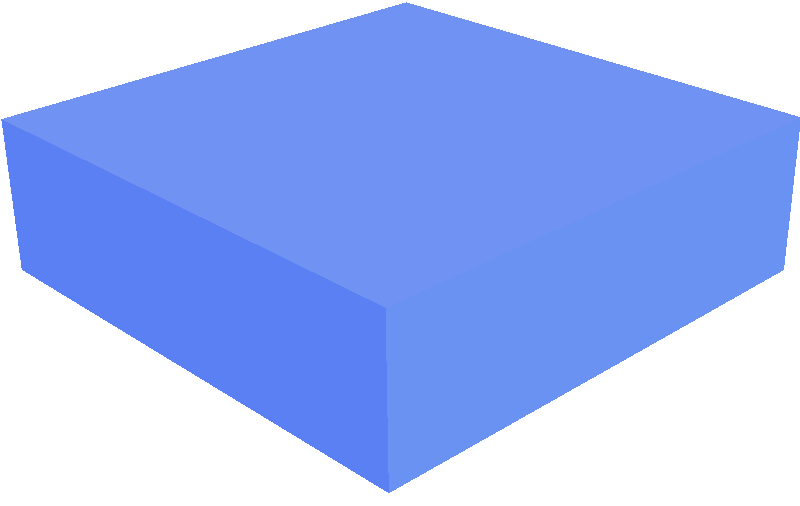}\\
\includegraphics[width=0.3\textwidth]{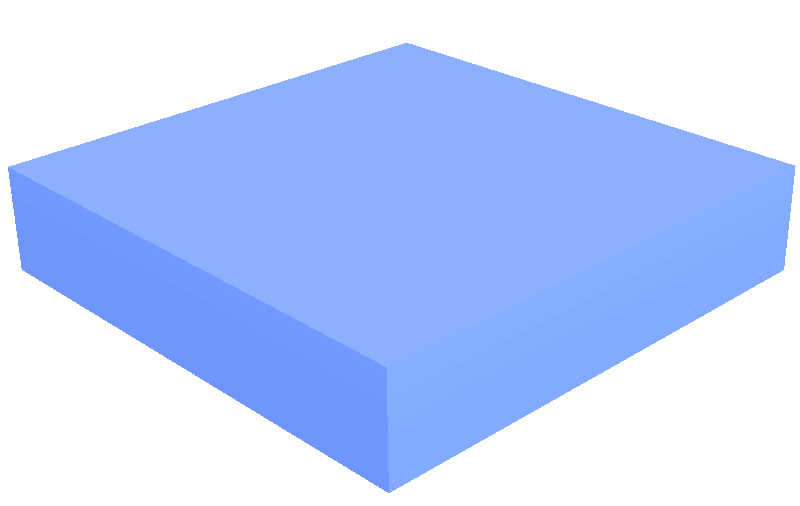}\\
\includegraphics[width=0.3\textwidth]{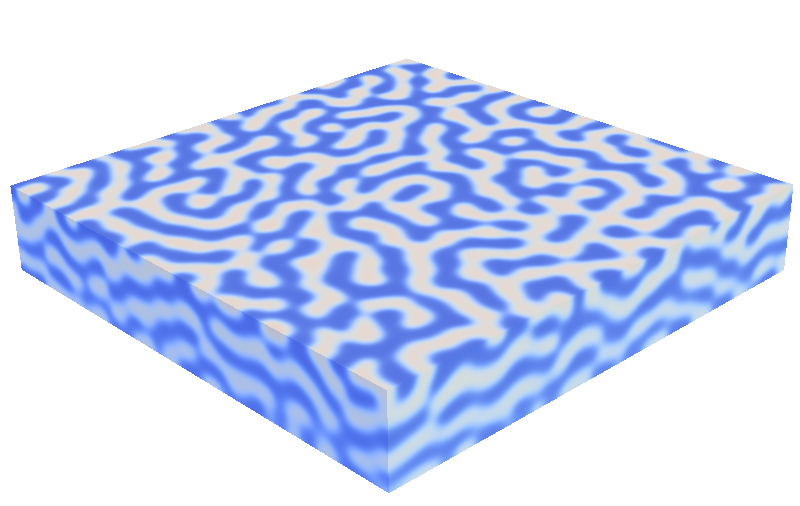}\\
\includegraphics[width=0.3\textwidth]{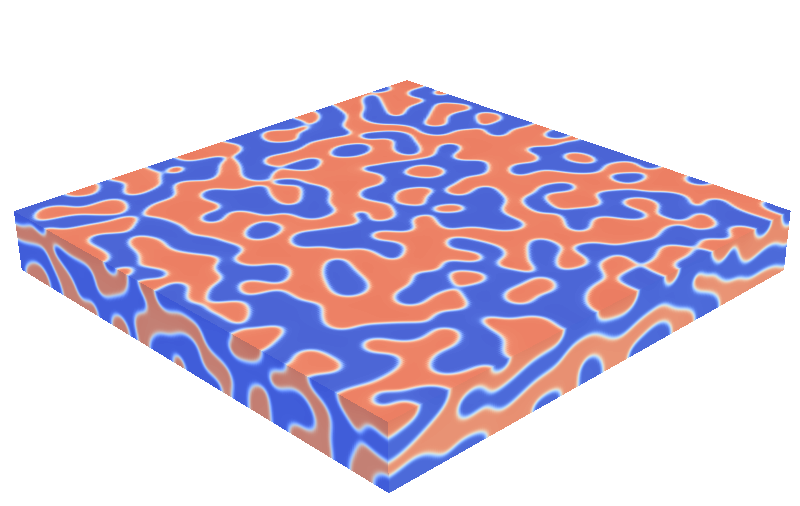}\\
\includegraphics[width=0.3\textwidth]{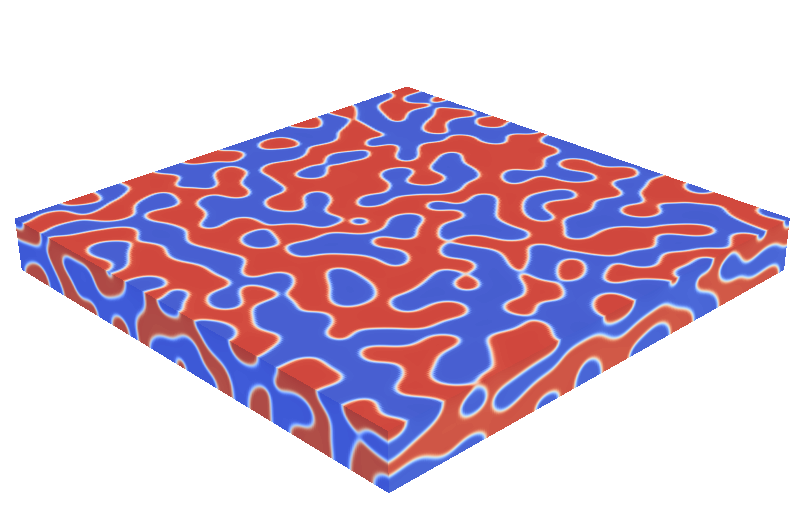}
}
\parbox{0.31\textwidth}{\centering $Bi=3$\\
\includegraphics[width=0.3\textwidth]{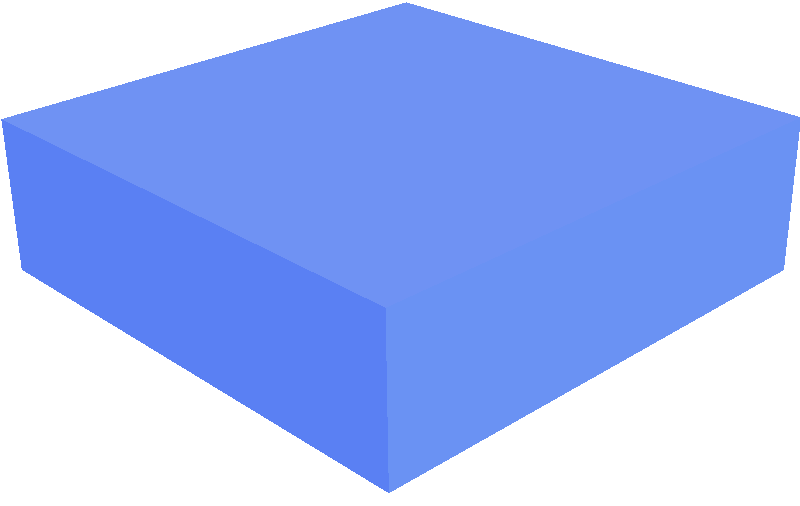}\\
\includegraphics[width=0.3\textwidth]{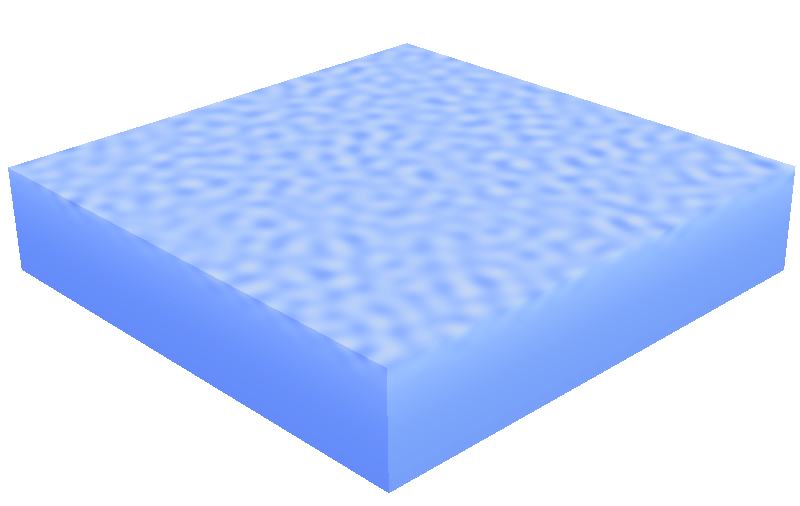}\\
\includegraphics[width=0.3\textwidth]{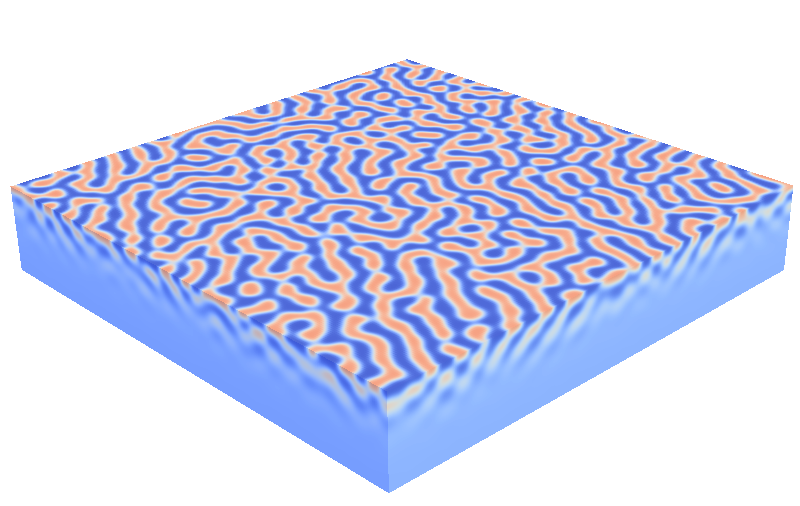}\\
\includegraphics[width=0.3\textwidth]{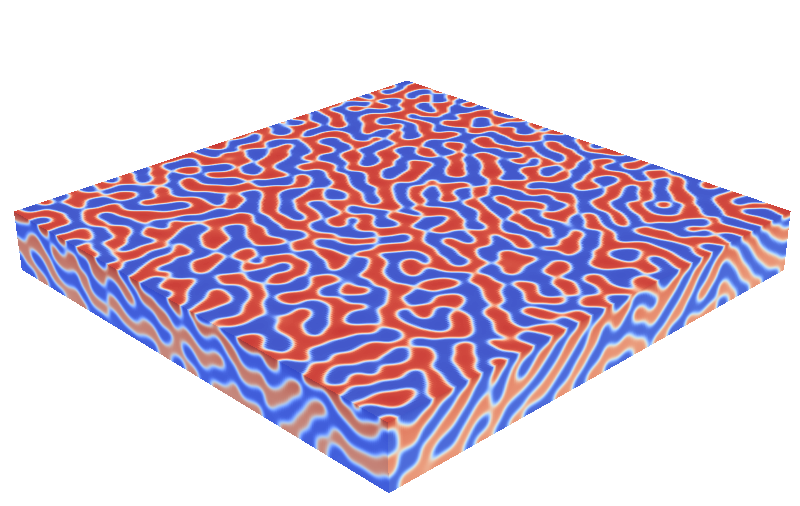}\\
\includegraphics[width=0.3\textwidth]{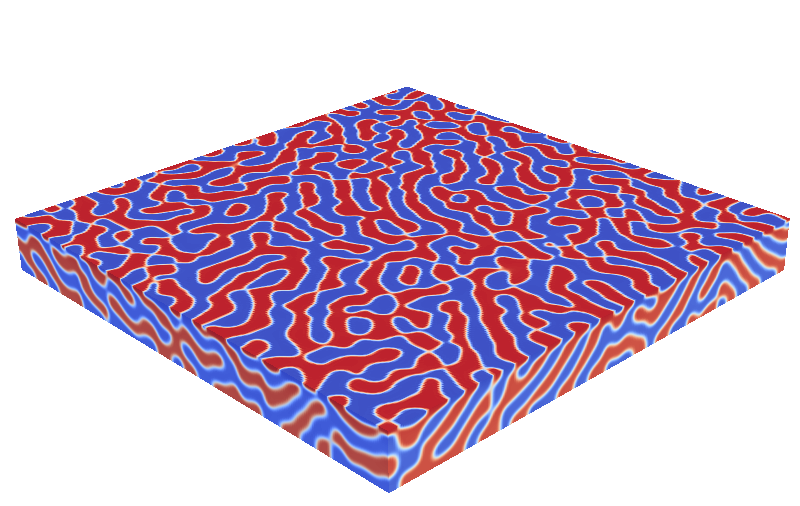}
}
\caption{Morphology evolution for three different Biot numbers: $Bi=0.03$, $0.3$ and $3$. Consecutive rows correspond to the morphology at height: initial, $h=0.7$, $0.6$, $0.4$ and the final height.}
\label{fig:res:Bi:3D}
\end{figure}

The evaporation rate affects not only the region where phase separation is initiated but also affects the average size of the separated phases. 
In Figure~\ref{fig:res:Bi:3D}, we plot morphology evolution for three different evaporation rates characterized by Biot number $0.03$, $0.3$ and~$3$. The magnitude of the evaporation rate affects the total time of the process (as shown in Figure~\ref{fig:res:Bi}). 
For lower evaporation rate, and Biot number, total process time is longer. 
Once phases separates, domains rich in each component have additional time to coarsen and create larger domain (see Figure~\ref{fig:res:Bi:3D} left). 
For higher evaporation rate, and Biot number, total time is shorter. 
Once phases separates they have very short time to coarsen (see Figure~\ref{fig:res:Bi:3D} right). 
This is due to the fact that solvent is removed from the system rapidly, and consequently the diffusion coefficient of the film decreases significantly leaving the morphology frozen.

\subsection{Influence of blend ratio}

The blend ratio between polymer and fullerene is a key system variable during fabrication of OSC, that is additionally relatively easy to manipulate. The blend ratio affects the type of morphology that develops during spinodal decomposition.
Two basic classes of morphologies typically develop --- percolated morphology and morphology with islands.  Intuitively, the former is more suitable for OSC than the latter. 
This is because the latter structure has more islands not connected to relevant electrodes and cannot provide useful pathways for charges to reach the boundaries. 
Therefore, for OSC application, it is necessary to clearly identify process and system variables that lead to the percolated type of structure. 

\begin{figure}
\begin{flushright}
\includegraphics[width=0.15\textwidth]{3D_legend.pdf}\\
\end{flushright}
\centering
\parbox{0.05\textwidth}{
\setlength{\unitlength}{2mm}
\begin{picture}(10,80)
\put(0,50){$t$}
\thicklines
\put(3,65){\vector(0,-1){50}}
\end{picture}
}
\parbox{0.32\textwidth}{
\centering 1:1 \\
\includegraphics[width=0.3\textwidth]{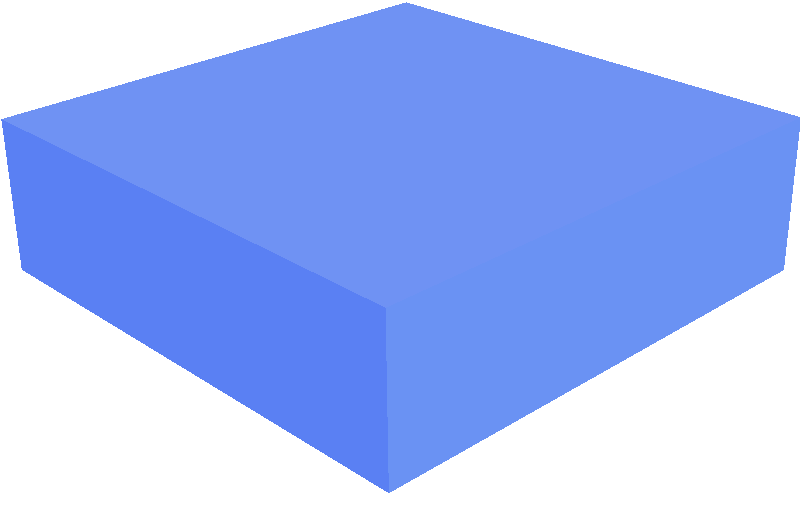}
\includegraphics[width=0.3\textwidth]{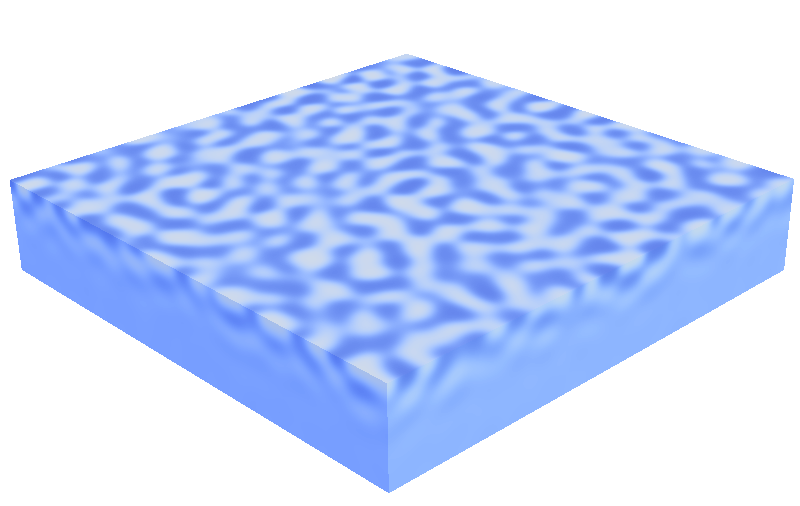}
\includegraphics[width=0.3\textwidth]{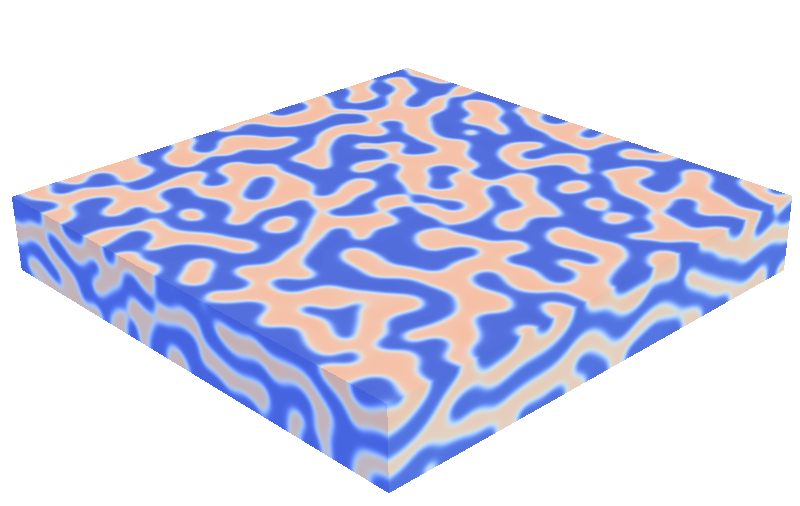}
\includegraphics[width=0.3\textwidth]{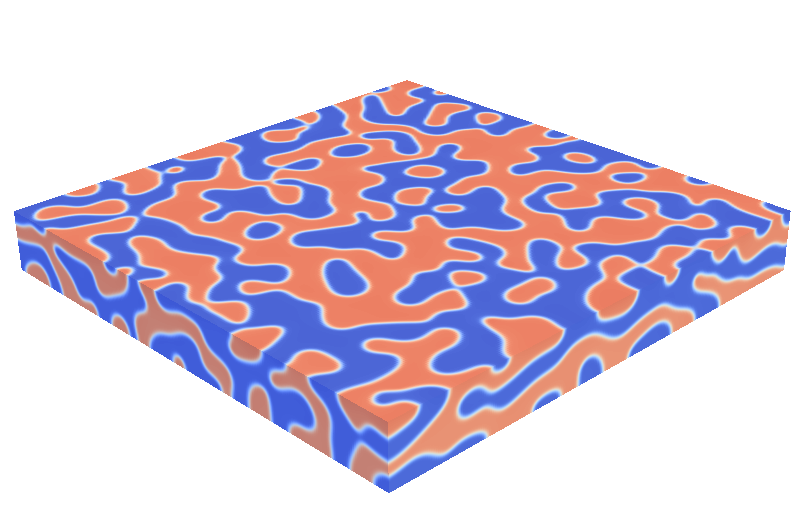}
\includegraphics[width=0.3\textwidth]{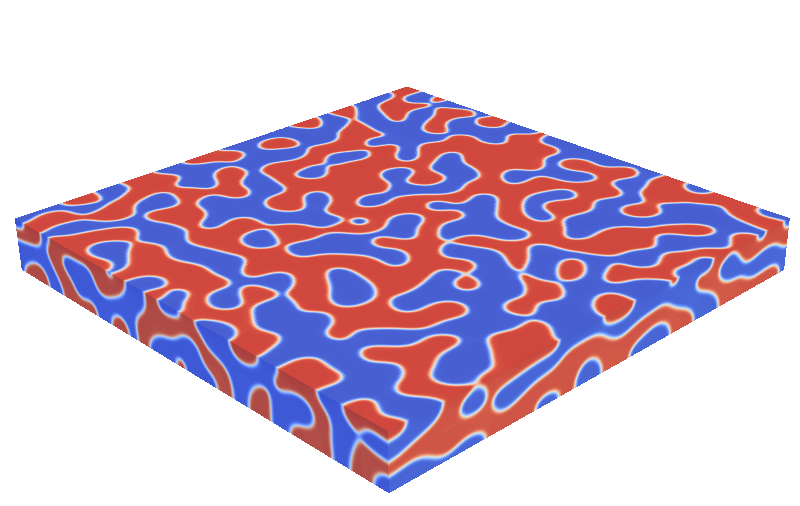}
}
\parbox{0.32\textwidth}{
\centering 1:0.8\\
\includegraphics[width=0.3\textwidth]{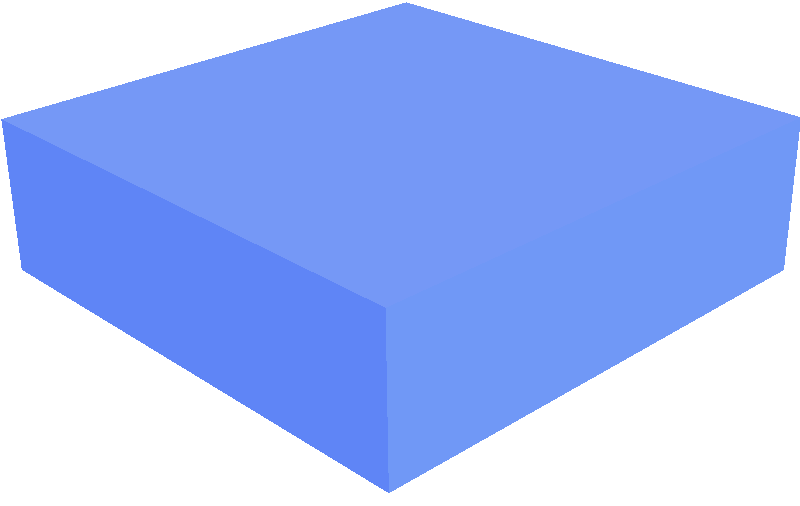}
\includegraphics[width=0.3\textwidth]{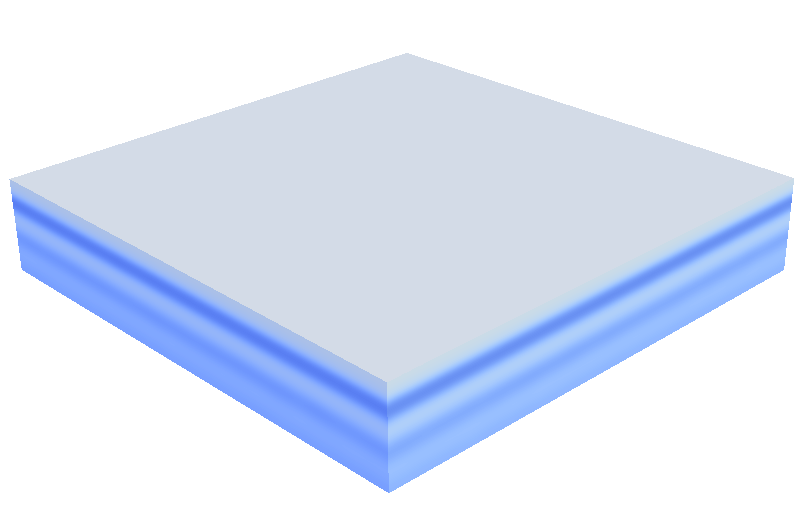}
\includegraphics[width=0.3\textwidth]{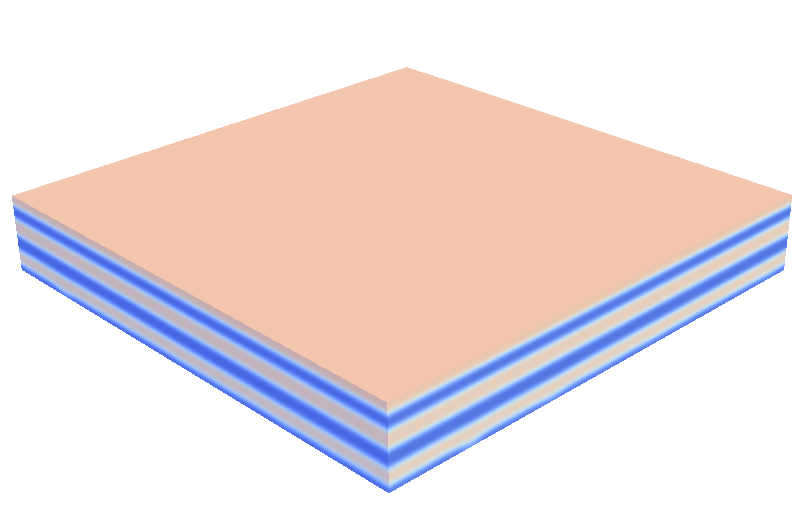}
\includegraphics[width=0.3\textwidth]{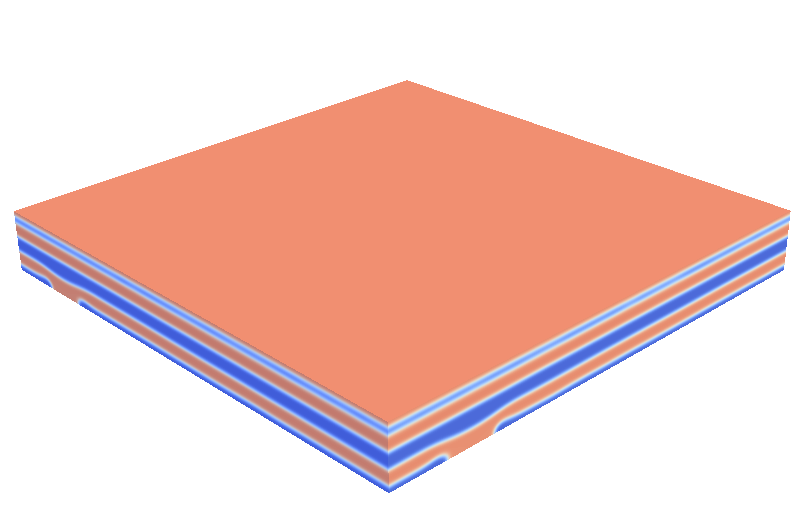}
\includegraphics[width=0.3\textwidth]{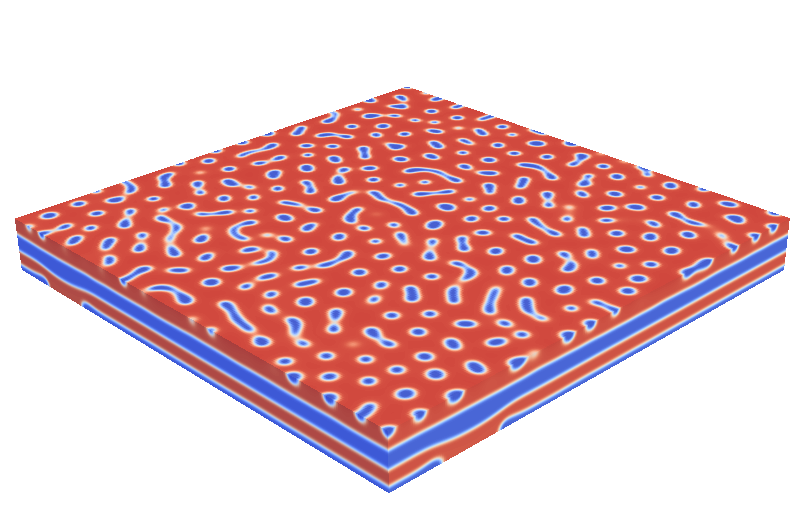}
}
\caption{Morphology evolution for two different blend ratio between polymer and fullerene 1:1 and 1:0.8. Consecutive rows correspond to the morphology at height: initial, $0.6$, $0.5$, $0.4$ and the final height.}
\label{fig:res:3D:BR}
\end{figure}

We investigate two blend ratios 1:1 and 1:0.8. 
In Figure~\ref{fig:res:3D:BR}, we plot snapshots of the morphology evolution with time.
Percolated morphology develops for blend ratio 1:1. 
However, a small change in blend ratio of the analyzed system results in significantly different morphology type.  This is clearly seen for blend ratio 1:0.8, where multiple stripes form spontaneously and break up as the solvent evaporates. Such multiple layer formation was also observed in experiments and is reported~\cite{BjorstromMoon2005}.  We note that although a layered morphology has no application in OSCs, such morphology is desired in organic transistors~\cite{CroneLi2000}.

\subsection{Influence of degree of polymerization}
\label{subsec:N}
\begin{figure}
\begin{flushright}
\includegraphics[width=0.18\textwidth]{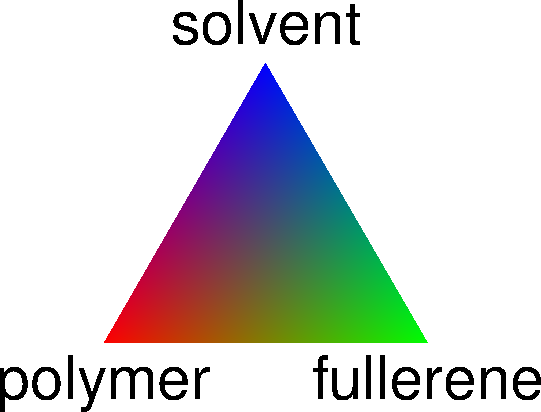}
\end{flushright}
\centering
\parbox{0.05\textwidth}{
\setlength{\unitlength}{2mm}
\begin{picture}(10,60)
\put(0,30){$t$}
\thicklines
\put(3,50){\vector(0,-1){40}}
\end{picture}
}
\parbox{0.3\textwidth}{
\centering$N_p=5$, $N_f=5$, $N_s=1$\\
\includegraphics[width=0.28\textwidth]{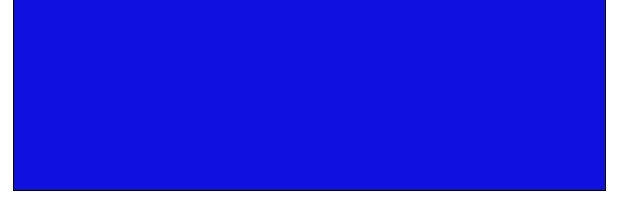}
\includegraphics[width=0.28\textwidth]{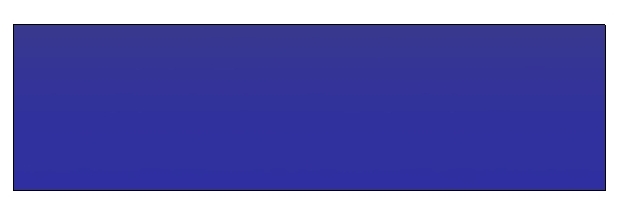}
\includegraphics[width=0.28\textwidth]{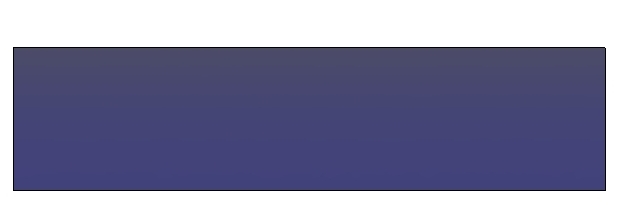}
\includegraphics[width=0.28\textwidth]{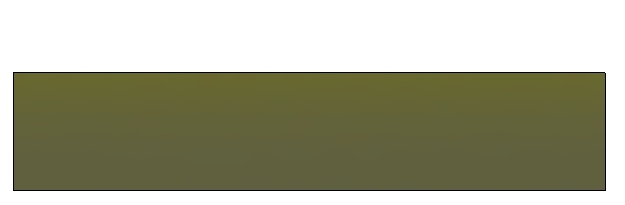}
\includegraphics[width=0.28\textwidth]{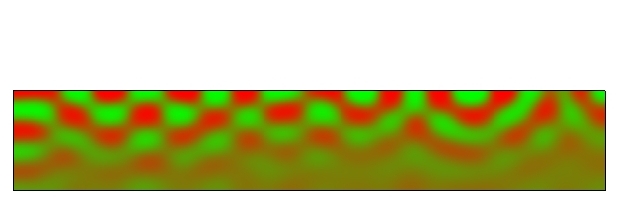}
\includegraphics[width=0.28\textwidth]{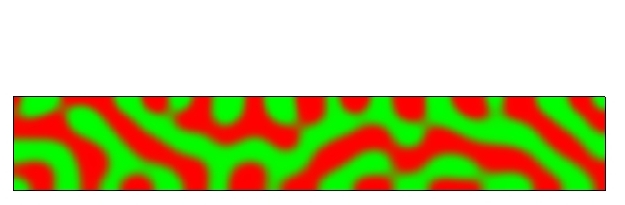}
\includegraphics[width=0.28\textwidth]{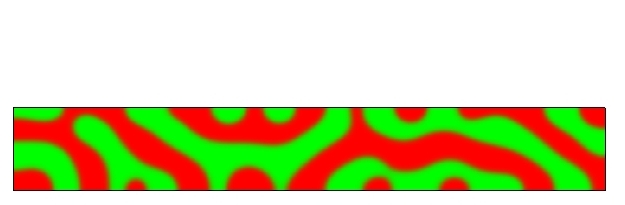}
\includegraphics[width=0.28\textwidth]{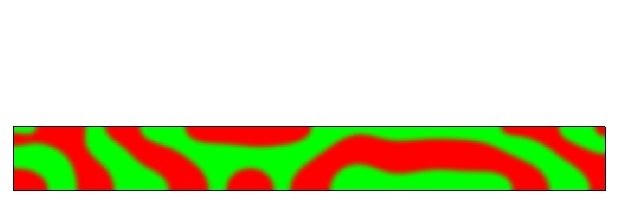}
}
\parbox{0.3\textwidth}{
\centering$N_p=20$, $N_f=5$, $N_s=1$\\
\includegraphics[width=0.28\textwidth]{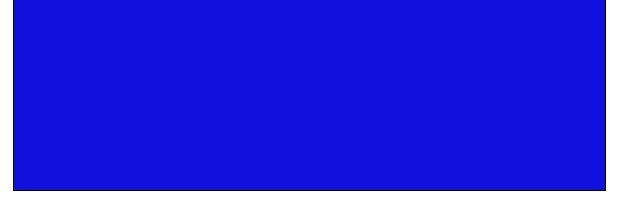}
\includegraphics[width=0.28\textwidth]{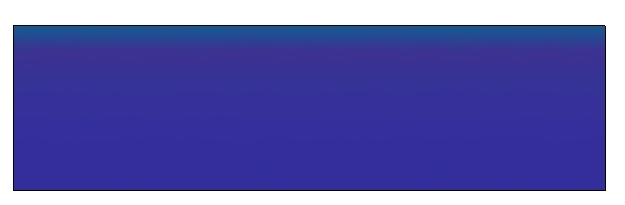}
\includegraphics[width=0.28\textwidth]{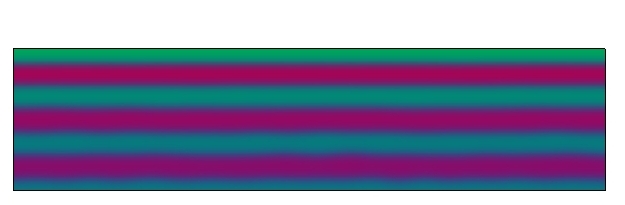}
\includegraphics[width=0.28\textwidth]{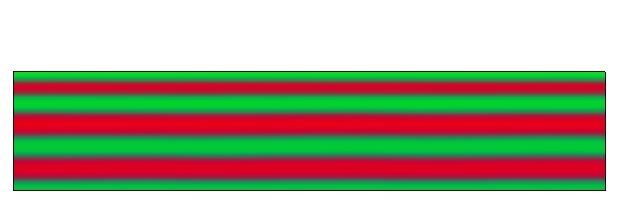}
\includegraphics[width=0.28\textwidth]{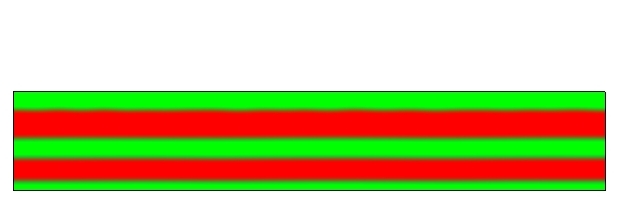}
\includegraphics[width=0.28\textwidth]{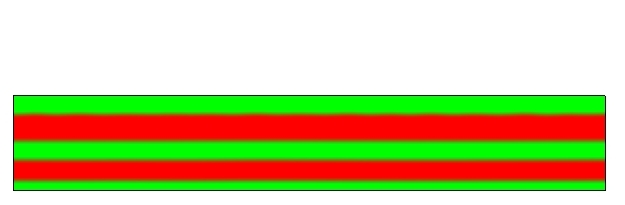}
\includegraphics[width=0.28\textwidth]{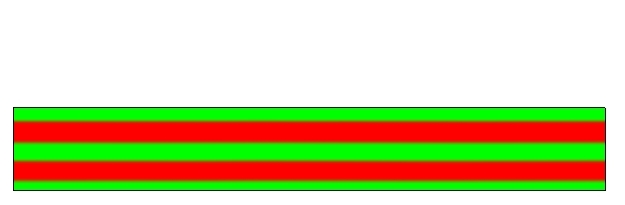}
\includegraphics[width=0.28\textwidth]{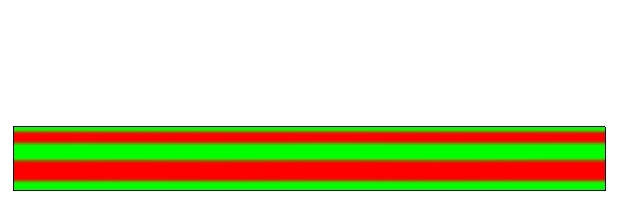}
}
\parbox{0.3\textwidth}{
\centering$N_p=100$, $N_f=5$, $N_s=1$\\
\includegraphics[width=0.28\textwidth]{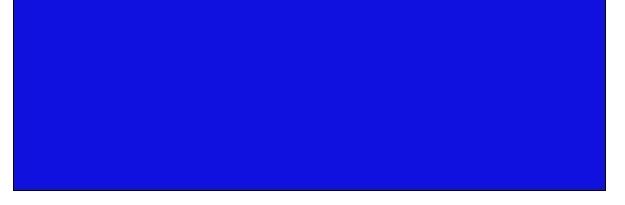}
\includegraphics[width=0.28\textwidth]{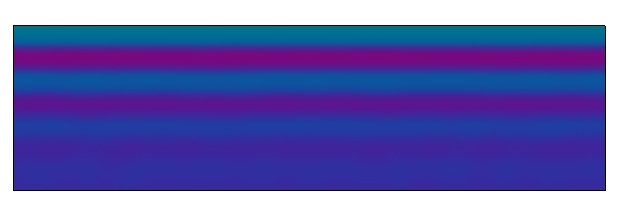}
\includegraphics[width=0.28\textwidth]{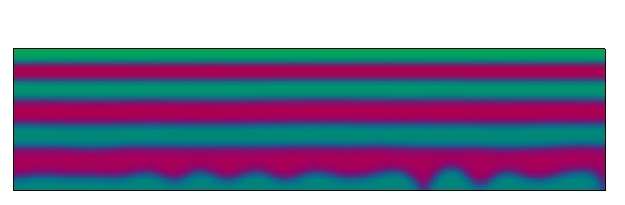}
\includegraphics[width=0.28\textwidth]{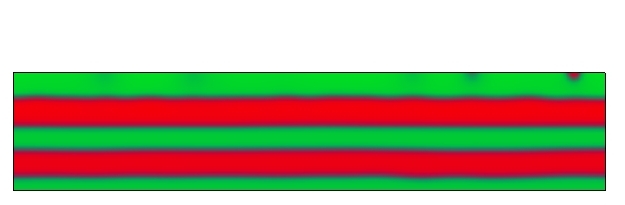}
\includegraphics[width=0.28\textwidth]{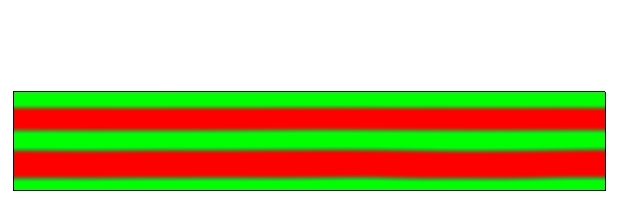}
\includegraphics[width=0.28\textwidth]{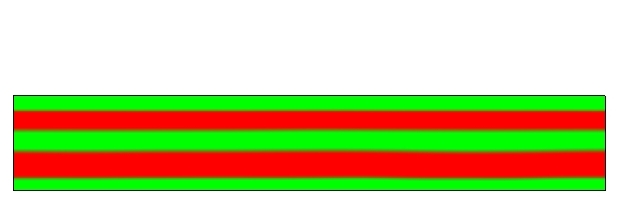}
\includegraphics[width=0.28\textwidth]{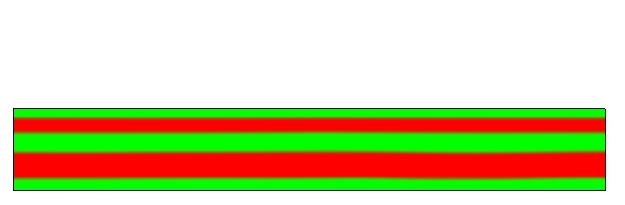}
\includegraphics[width=0.28\textwidth]{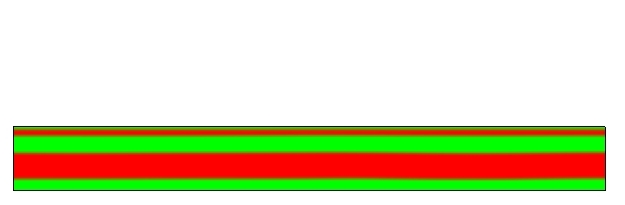}
}
\caption{Morphology evolution for three different degree of polymerization: $N_p=5$, $N_p=20$ and $N_p=100$. Consecutive rows correspond to the morphology at height: initial, $h=0.7$, $0.6$, $0.5$, $0.42$, $0.4$, $0.35$ and the final height.}
\label{fig:res:2D:N}
\end{figure}

The degree of polymerization, $N_p$, is another tunable variable that allows for controlling morphology evolution. In practice, for OSC fabrication, only polymer degree of polymerization can be controlled and it has been shown experimentally that efficiency is affected by its choice~\cite{VeenstraKroon2007}. In Figure~\ref{fig:res:2D:N}, we compare the morphology evolution for three different degree of polymerization: $N_p=5$, $20$ and $100$. 
All simulations have been performed for the same blend ratio 1:1 and Biot number $Bi=0.4$.
As we increase the degree of polymerization, the morphology changes significantly. 
For the symmetric case: $N_p=N_f=5$, a percolated morphology evolves; while for asymmetric cases multiple layers are created. Notice also that with increasing degree of polymerization,  phase separation is initiated earlier.  Moreover, when polymer degree of polymerization is larger than fullerene degree of polymerization, morphology type changes from percolated into multiple layered morphology. 

\subsection{Influence of solvent type}
\label{subsec:solvent}
In solvent-based fabrication, the solvent creates the environment for morphology evolution. 
Both components must be soluble in the common solvent to create an initially homogeneous solution. Choice of the solvent has significant effect on morphology evolution and provides additional system variable for morphology control. This is manifested as differences in relative values of interaction parameters between all three components. 

\begin{figure}
\begin{flushright}
\includegraphics[width=0.18\textwidth]{RGB_legend.png}
\end{flushright}
\parbox{0.05\textwidth}{
\setlength{\unitlength}{2mm}
\begin{picture}(10,60)
\put(0,30){$t$}
\thicklines
\put(3,50){\vector(0,-1){40}}
\end{picture}
}
\parbox{0.3\textwidth}{
\centering$\chi_{ps}=0.3$,$\chi_{fs}=0.3$\\
\includegraphics[width=0.28\textwidth]{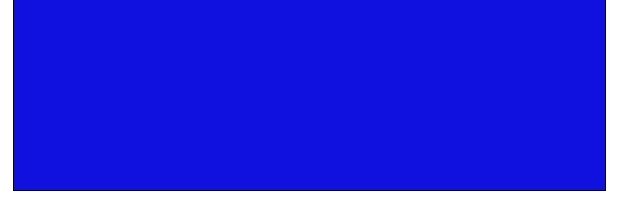}
\includegraphics[width=0.28\textwidth]{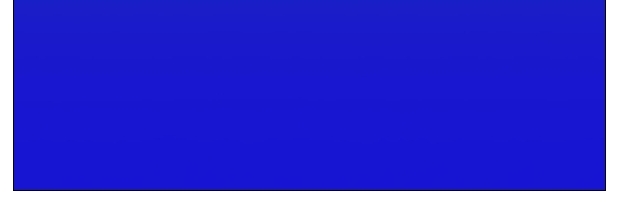}
\includegraphics[width=0.28\textwidth]{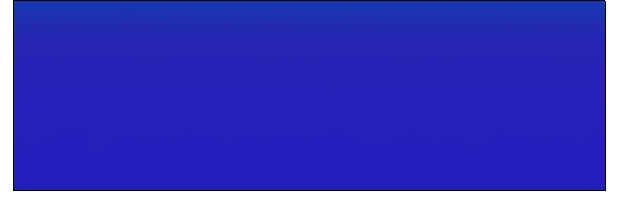}
\includegraphics[width=0.28\textwidth]{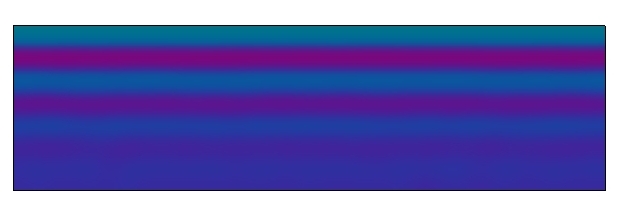}
\includegraphics[width=0.28\textwidth]{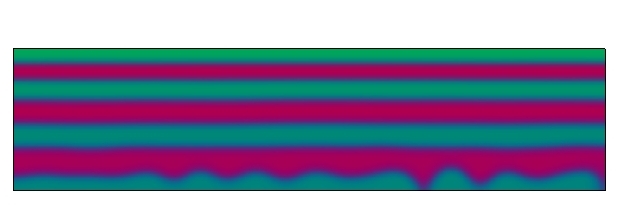}
\includegraphics[width=0.28\textwidth]{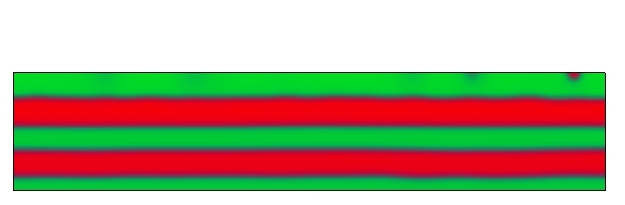}
\includegraphics[width=0.28\textwidth]{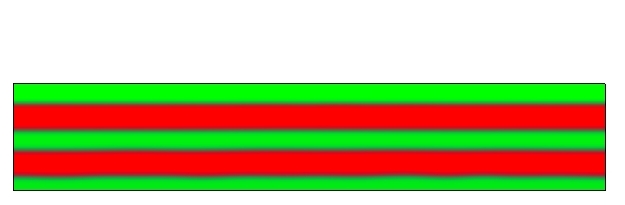}
\includegraphics[width=0.28\textwidth]{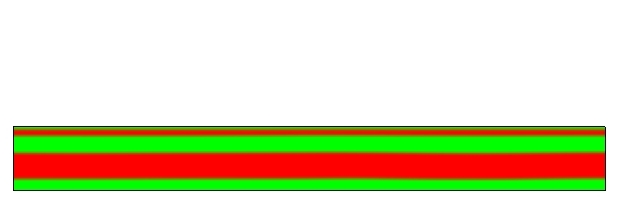}
}
\parbox{0.3\textwidth}{
\centering$\chi_{ps}=0.3$,$\chi_{fs}=0.6$\\
\includegraphics[width=0.28\textwidth]{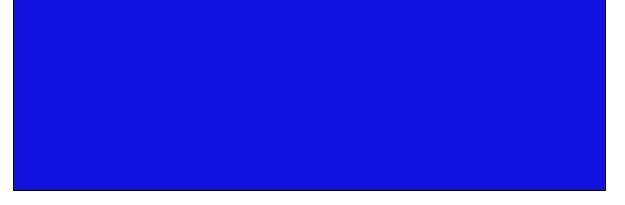}
\includegraphics[width=0.28\textwidth]{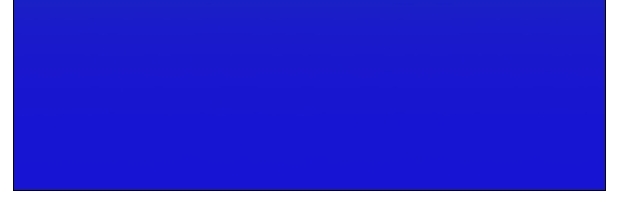}
\includegraphics[width=0.28\textwidth]{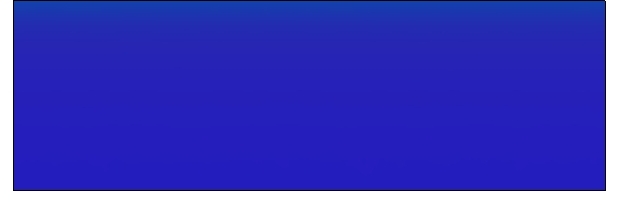}
\includegraphics[width=0.28\textwidth]{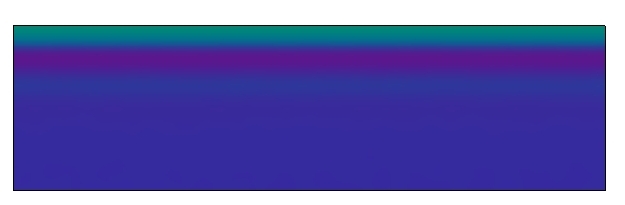}
\includegraphics[width=0.28\textwidth]{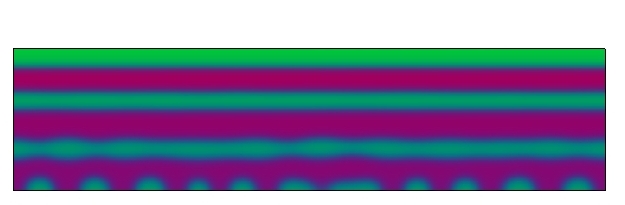}
\includegraphics[width=0.28\textwidth]{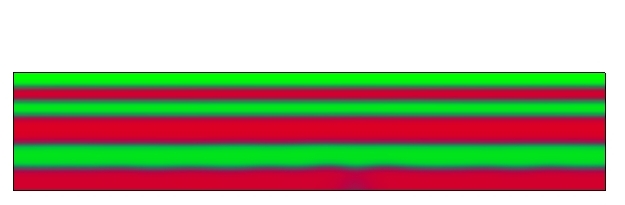}
\includegraphics[width=0.28\textwidth]{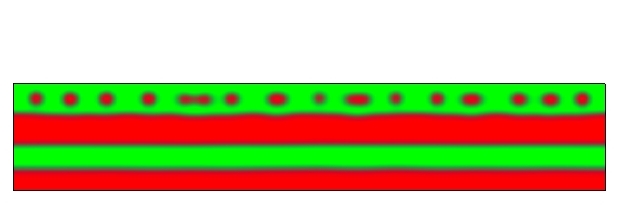}
\includegraphics[width=0.28\textwidth]{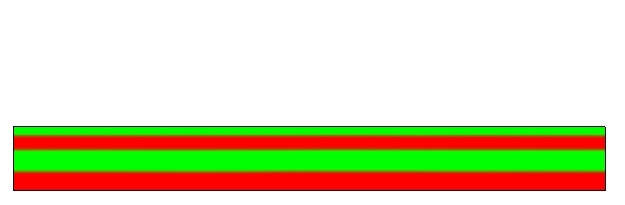}
}
\parbox{0.3\textwidth}{
\centering$\chi_{ps}=0.6$,$\chi_{fs}=0.3$\\
\includegraphics[width=0.28\textwidth]{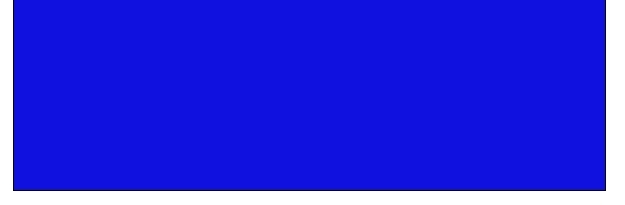}
\includegraphics[width=0.28\textwidth]{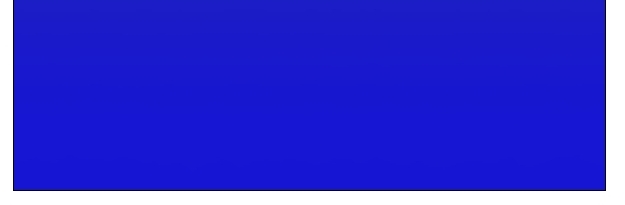}
\includegraphics[width=0.28\textwidth]{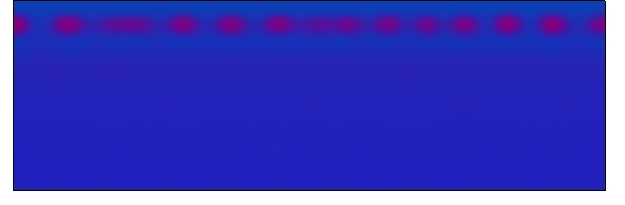}
\includegraphics[width=0.28\textwidth]{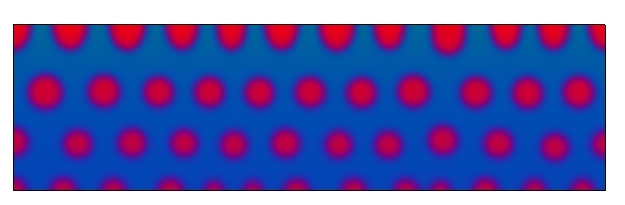}
\includegraphics[width=0.28\textwidth]{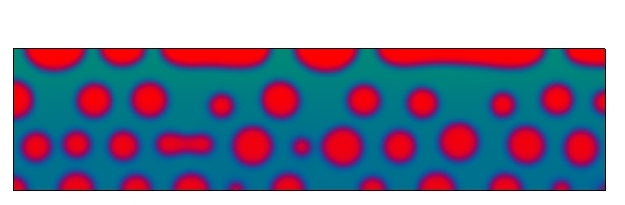}
\includegraphics[width=0.28\textwidth]{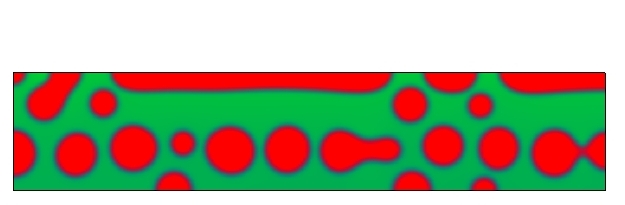}
\includegraphics[width=0.28\textwidth]{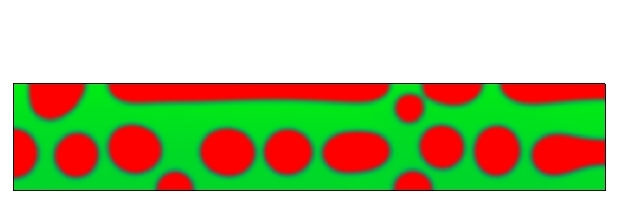}
\includegraphics[width=0.28\textwidth]{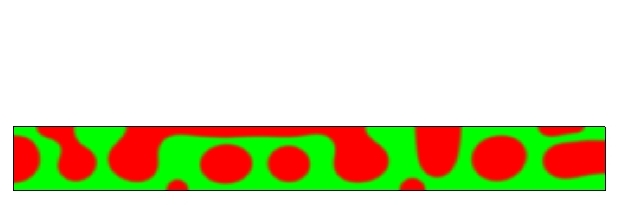}
}
\caption{Morphology evolution for three different solvent types characterized by different interaction parameters between components. Consecutive rows correspond to the morphology at height: initial, $h=0.9$, $0.8$, $0.7$, $0.6$, $0.5$, $0.45$ and the final height.}
\label{fig:res:2D:solvent}
\end{figure}

In Figure~\ref{fig:res:2D:solvent}, we show morphology evolution for three combinations of the interaction parameters ($\chi_{pf}$, $\chi_{ps}$, $\chi_{fs}$): ($1$,$0.3$, $0.3$), ($1$,$0.3$, $0.6$) and ($1$,$0.6$, $0.3$).
In the first case, solvent is chosen such that interactions between solvent and two components are the same and much lower than interaction between polymer and fullerene. This means that polymer and fullerene have similar solubilities in solvent. In the second case, fullerene is less soluble in the solvent compared to polymer. Interaction parameter between fullerene and solvent is two times higher than between polymer and solvent. In the third case, we assume that polymer is less soluble in the solvent.  In all three case, we assume the same evaporation rate ($Bi=0.3$), blend ratio (1:1), and  degree of polymerization ($N_p=100$, $N_{f}=5$ and $N_s=1$).

Changing solvent results in dramatically different morphology evolution~\cite{SBS01}. Multiple layer formation that we observe for higher polymerization is broken in the third case, due to the different solvent used. 

\subsection{Effect of substrate}
The height of a typical OSC active layer is around $100 - 300\;nm$. 
For such thin geometries, morphology close to the substrate can be significantly different than that of the bulk. We consider the effect of chemical patterning of substrate to selectively affect each component. We consider two cases. In the first case, the substrate is patterned to attract the polymer preferentially. In the second case, substrate is patterned with two chemistries: one preferentially attracting polymer and one preferentially attracting fullerene. In both cases, patterns of wavelength $\lambda_s = 0.12$ cover the substrate.
In each case, we assume the chemical potential of patterned material $\mu^p=\mu^f=0.01$, and $h^p=h^f=0$. In most cases, $h$ is small compared to $\mu$~\cite{J93}.

\begin{figure}
\begin{flushright}
\includegraphics[width=0.18\textwidth]{RGB_legend.png}
\end{flushright}
\parbox{0.05\textwidth}{
\setlength{\unitlength}{2mm}
\begin{picture}(10,60)
\put(0,30){$t$}
\thicklines
\put(3,50){\vector(0,-1){40}}
\end{picture}
}
\centering
\parbox{0.3\textwidth}{
\centering no\\ substrate patterning\\
\includegraphics[width=0.28\textwidth]{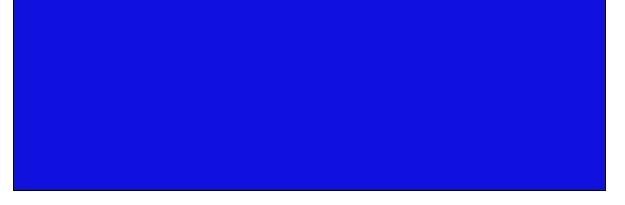}
\includegraphics[width=0.28\textwidth]{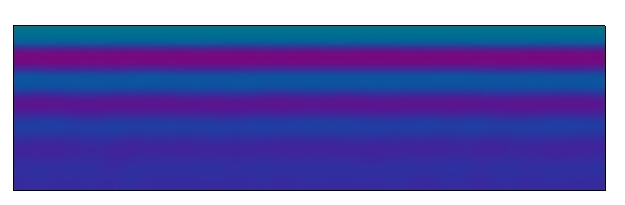}
\includegraphics[width=0.28\textwidth]{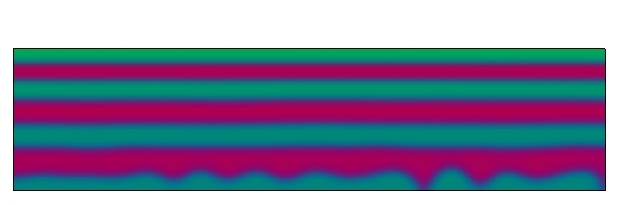}
\includegraphics[width=0.28\textwidth]{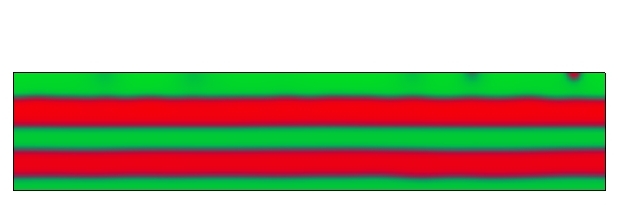}
\includegraphics[width=0.28\textwidth]{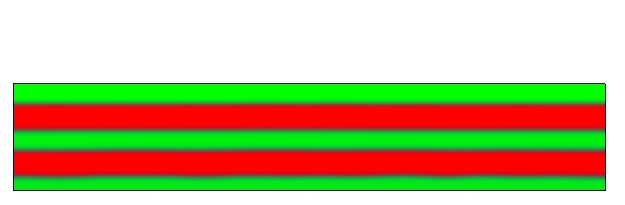}
\includegraphics[width=0.28\textwidth]{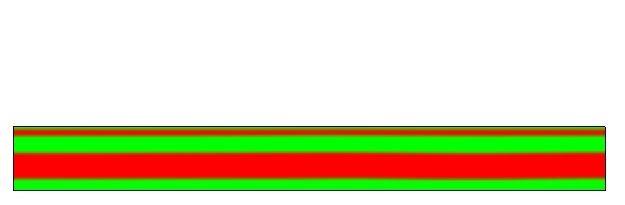}
}
\parbox{0.3\textwidth}{
\centering substrate patterning with one material\\
\includegraphics[width=0.28\textwidth]{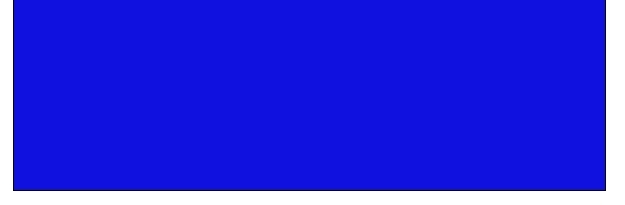}
\includegraphics[width=0.28\textwidth]{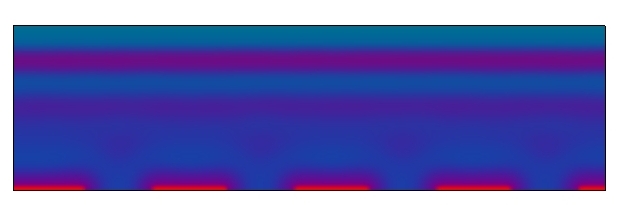}
\includegraphics[width=0.28\textwidth]{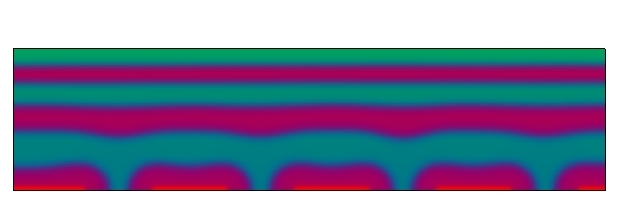}
\includegraphics[width=0.28\textwidth]{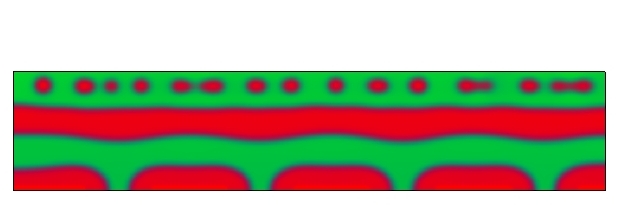}
\includegraphics[width=0.28\textwidth]{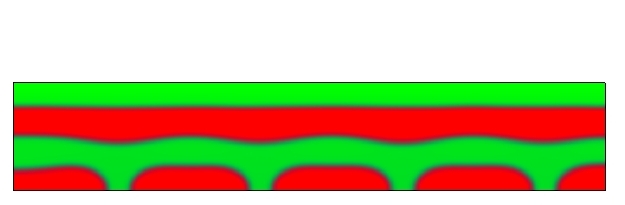}
\includegraphics[width=0.28\textwidth]{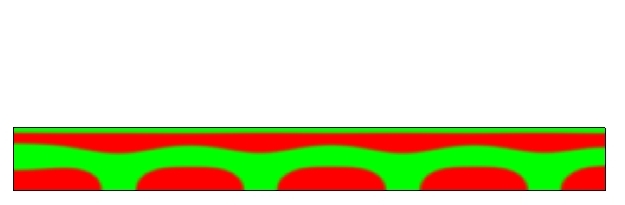}
}
\parbox{0.3\textwidth}{
\centering substrate patterning with two materials\\
\includegraphics[width=0.28\textwidth]{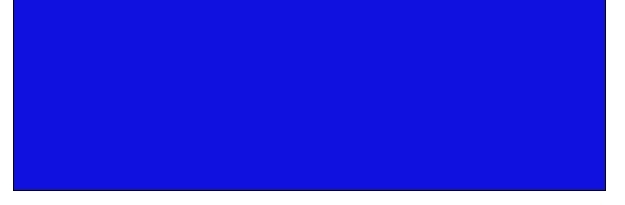}
\includegraphics[width=0.28\textwidth]{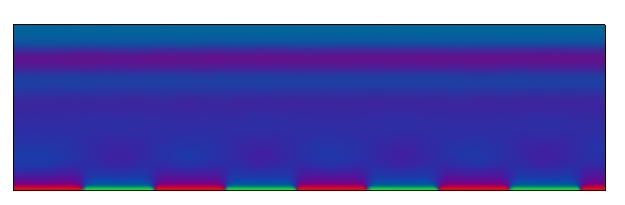}
\includegraphics[width=0.28\textwidth]{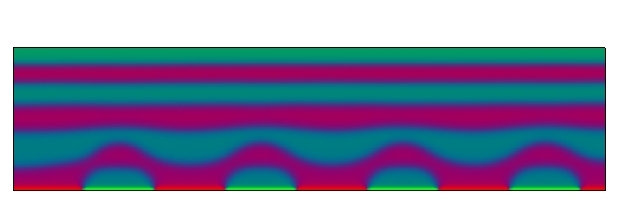}
\includegraphics[width=0.28\textwidth]{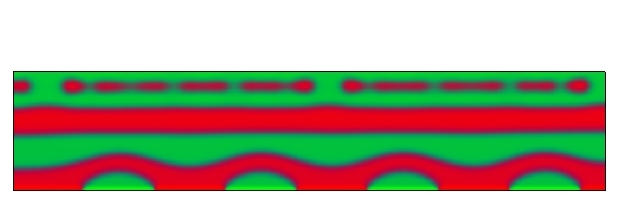}
\includegraphics[width=0.28\textwidth]{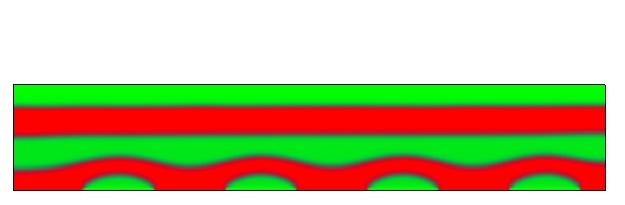}
\includegraphics[width=0.28\textwidth]{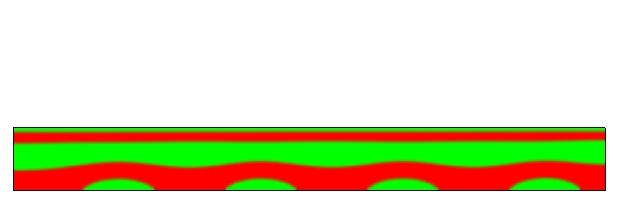}
}
\caption{Morphology evolution three systems where substrate is neutral to both components (left), substrate is patterned with material that attracts polymer (middle) and substrate is patterned with two materials, one attracting polymer, second attracting fullerene (right). Consecutive rows presents morphology evolution at $h=1$, $0.7$, $0.6$, $0.5$, $0.45$ and final.}
\label{fig:res:2D:substrate}
\end{figure}

Figure~\ref{fig:res:2D:substrate} shows morphology evolution on three different patterned substrates: neutral, preferentially attracting polymer, and preferentially attracting polymer and fullerene. We run these tests for polymer of high degree of polymerization $N_p=100$ (last column in Figure~\ref{fig:res:2D:N}~right), which becomes the reference simulation and is repeated in Figure~\ref{fig:res:2D:substrate} (left). 
\emph{These results clearly demonstrate that substrate patterning provides additional degree of control over morphology.} Multiple layers observed without patterning can be broken close to the substrate. The breakage depends on the frequency of the patterning and combination of the material types (Figure~\ref{fig:res:2D:substrate} middle and right). 

Patterning with alternating chemical preference allows for better control of domain size. 
The size along the substrate  of the polymer-rich induced by such patterning is maintained during evaporation. This is not the case when patterning is purely of one preference.  The domains created close to the substrate grow in size along the substrate.  However, when substrate is patterned with one chemical preference, domains created close to the substrate penetrate deeper into film, compared to the other case. 

\subsection{Control of morphology evolution}
\begin{figure}
\parbox{0.32\textwidth}{\centering a
\includegraphics[width=0.28\textwidth]{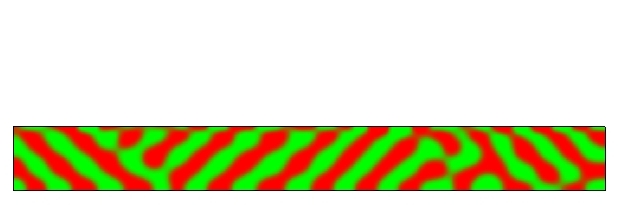}
}
\parbox{0.32\textwidth}{\centering b
\includegraphics[width=0.28\textwidth]{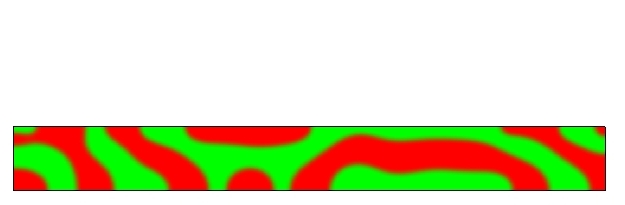}
}
\parbox{0.32\textwidth}{\centering c
\includegraphics[width=0.28\textwidth]{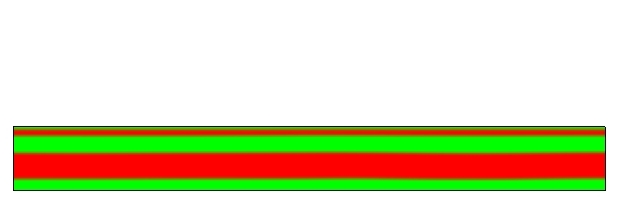}
}\\
\parbox{0.32\textwidth}{\centering d
\includegraphics[width=0.28\textwidth]{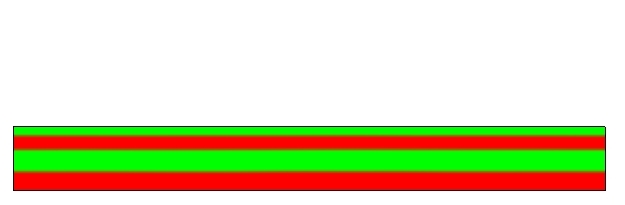}
}
\parbox{0.32\textwidth}{\centering e
\includegraphics[width=0.28\textwidth]{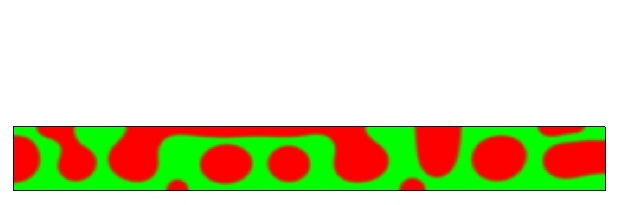}
}
\parbox{0.32\textwidth}{\centering f
\includegraphics[width=0.28\textwidth]{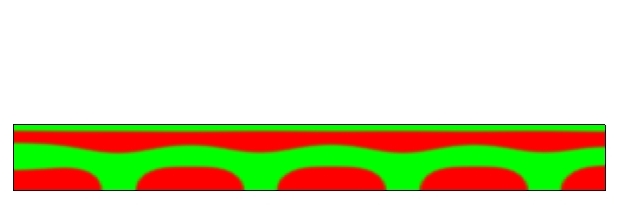}
}
\caption{Morphology control by changing system and process variables of the solvent-based fabrication techniques.}
\label{fig:res:2D:sum} 
\end{figure}

In previous subsections, we showed that by independently changing system and process variables we obtain various types of morphologies. 
In Figure~\ref{fig:res:2D:sum}, we summarize these types of morphologies.
It is important to notice that significantly different morphologies can develop for various system and process variables. 
In general, there are three main types of morphologies: percolated morphology (b); morphology with multiple layers~(c and d) and morphology with islands~(e). 
Substrate patterning gives an additional means to initiate and direct morphology close to the substrate. For example, adding substrate patterning leads to breaking the bottom layer, as shown in Figure~\ref{fig:res:2D:substrate}. 
{\it It is interesting to note that controlling different variables may lead to the same type of morphology.} For example, increasing degree of polymerization from $N_p=5$ to $N_p=20$ leads to multiple layer creation (Figure~\ref{fig:res:2D:N}). Similar effect is observed by changed solvent (Figure~\ref{fig:res:2D:solvent}). This emphasize the importance of further systematic studies of solvent-based fabrication.

\section{Conclusions}
\label{ch:C}
Morphology is a key element affecting the performance of organic solar cells. The morphology evolution during solvent-based fabrication of organic solar cells is a complex, multi physics process that is affected by a variety of material and process parameters. A virtual framework that can model 3-D morphology evolution during fabrication of OSC can relate these fabrication conditions with morphology. This will significantly augment organic photovoltaic research which has been predominantly based on experimental trial and error investigations. Such a framework will also allow high throughput analysis of the large phase space of processing parameters, thus yielding considerable insight into the process-structure-property relationships governing organic solar cell behavior that is currently in its infancy.

In this work, we develop a phase field-based framework to study 3D nanomorphology evolution in the active layer of OSC during solvent-based fabrication process. 
In particular, we formulate physical and mathematical model that takes into account all important processes that occur during solvent-based fabrication of OSCs. We select phase field method to model the behavior multicomponent system with various driving forces for morphology evolution. We outline an efficient numerical implementation of the framework, to enable three dimensional analysis of the process. We showcase our framework by investigating the effect of various process and system variables, that lead to following observations:
\begin{enumerate}
\item Mass Biot number expresses the interplay between solvent evaporation from the top surface and diffusion within the thin film. For high Biot number, evaporation is a dominant process which results in top boundary layer creation enriched in two main components. Consequently, phase separation initiates close to the top and propagates into the film. For low Biot number, in turn, diffusion is a dominant process, solvent removed from the top layer is diffused back from the resulting in homogeneous profile. Thanks to this uniform distribution phase separation initiates and evolve homogeneously within film. 
\item The morphology evolution is affected not only by kinetics through evaporation and diffusion but also by thermodynamics. In particular, interaction parameters between components and degree of polymerization have a large effect on morphology evolution.
\item The accessibility of possible configurations provided by the free energy landscape is controlled by system variable such as blend ratio. Small change of blend ratio lead to large variation in morphology evolution. 
\item Finally, surface induced phase separation provides another opportunity to locally affect the  morphology, by creating additional ''sinks`` in the energy landscape. 
\end{enumerate}
We are currently investigating extensions of the framework along three directions: nonhomogeneous evaporation, fluid shear effects (based on~\cite{GomezCalo2010}), and crystallization.

\section{Acknowledgments}
This research was supported in part by the National Science Foundation through TeraGrid resources provided by TACC under grant number TG-CTS110007 and TG-CTS100080. BG \& OW were supported in part by NSF PHY-0941576 and NSF CCF-0917202.

\appendix
\section{Time step adaptivity}
\label{sec:app}
\begin{figure}
\parbox{0.5\textwidth}{\centering 1D}
\parbox{0.5\textwidth}{\centering 2D}\\
\includegraphics[width=0.5\textwidth]{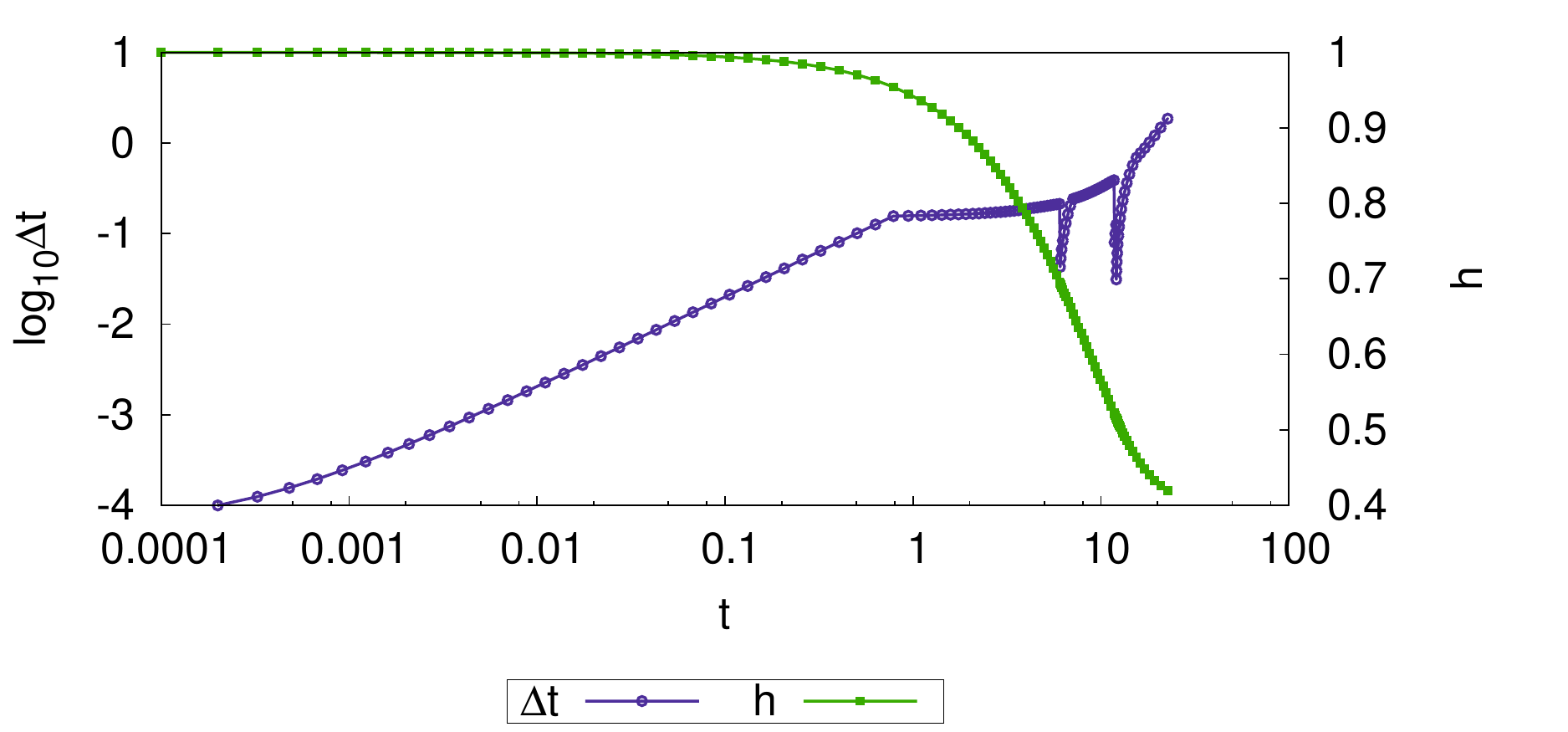}
\includegraphics[width=0.5\textwidth]{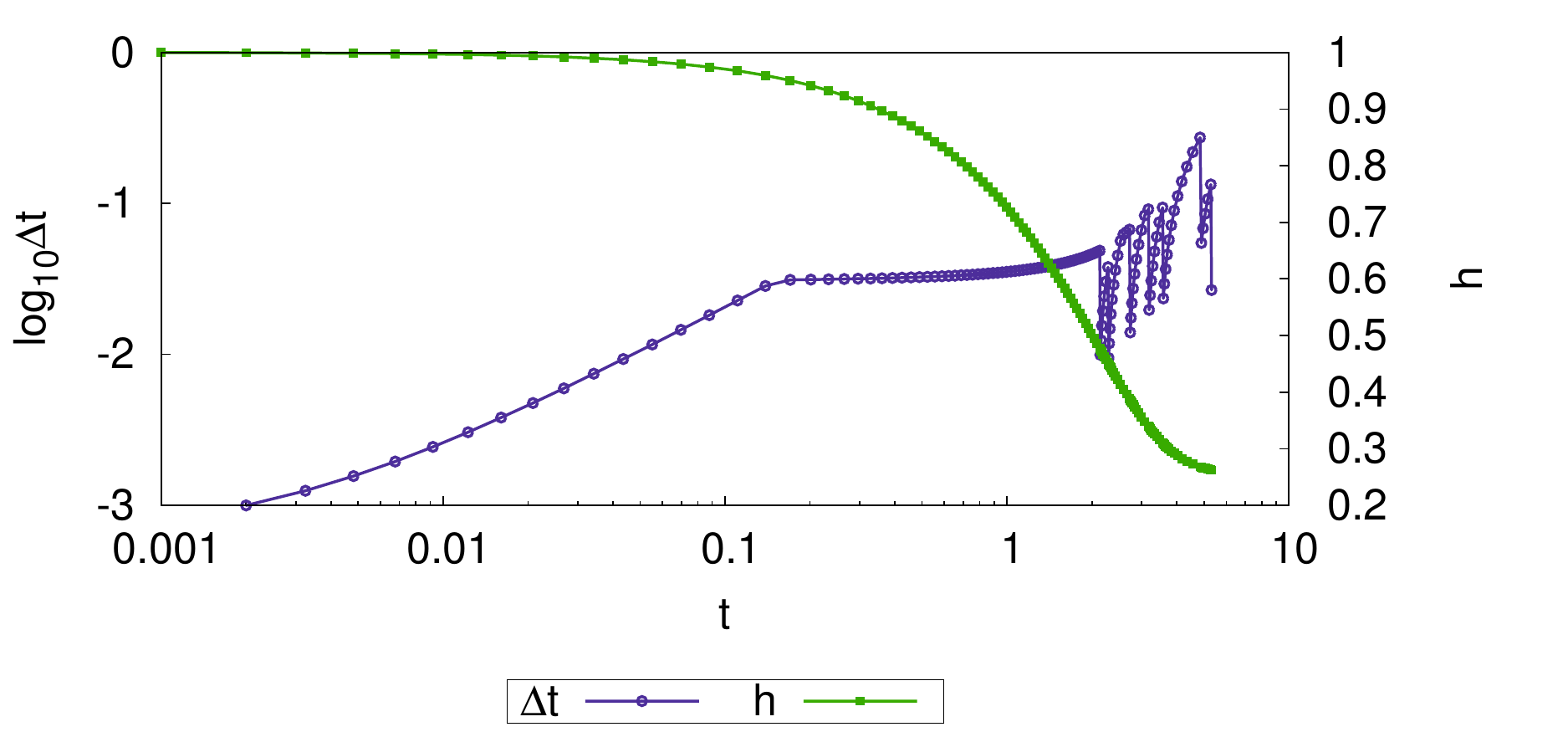}\\
\parbox{0.5\textwidth}{\centering 3D}\\
\includegraphics[width=0.5\textwidth]{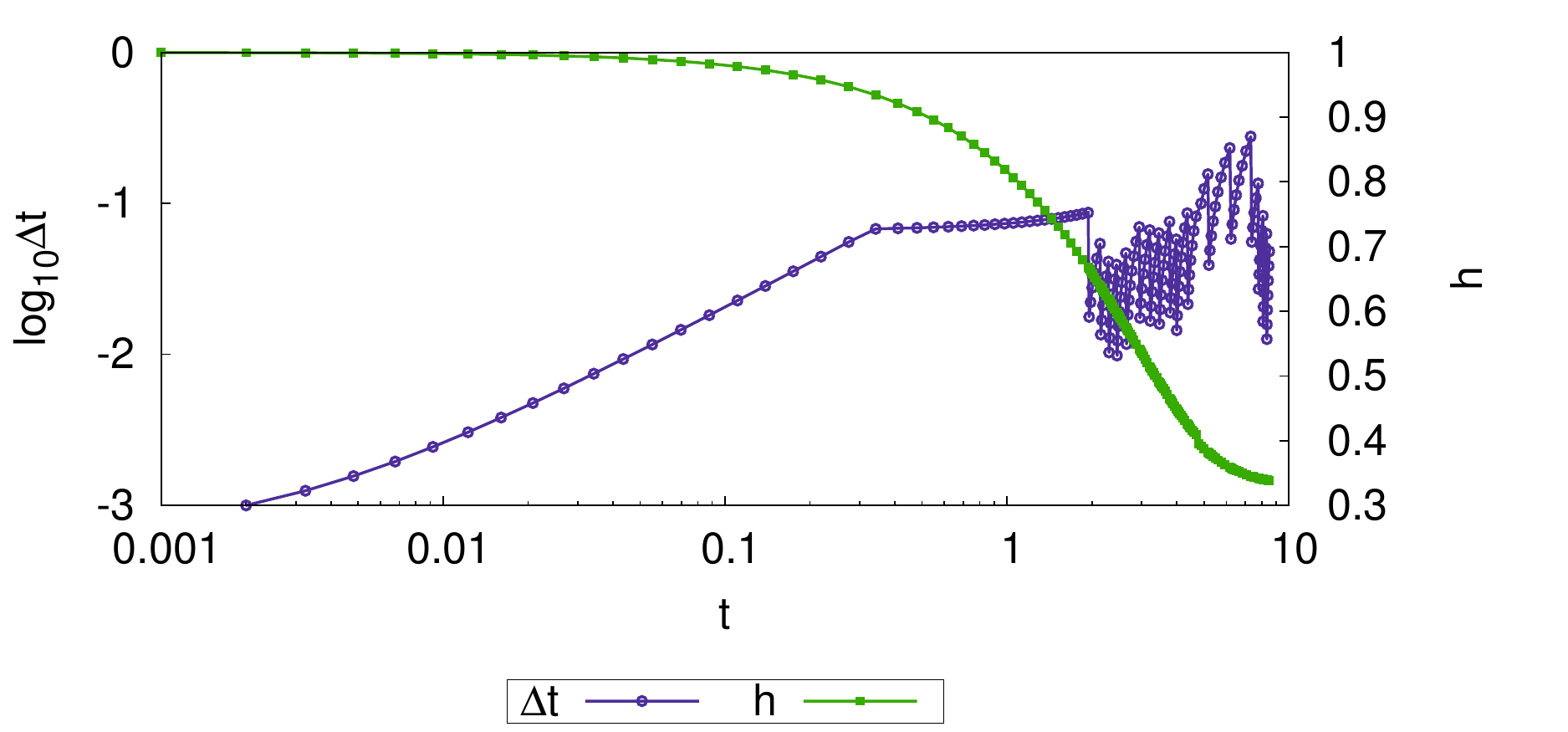}
\caption{Size of time step profile and film height profiles for example 1D, 2D, and 3D problems, respectively.}
\label{fig:res:ts}
\end{figure}

We use Euler Backward scheme with the heuristic strategy to adjust size of time step~\cite{RFKS10}. 
This strategy is based on the number of Newton's iterations required to solve a nonlinear problem for given time step. 
If number of iteration is lower than 20 size of time step is increased by 25\% in new time step, otherwise it is reduced to 25\% of previous time step. 
When solution cannot be found in 50 iterations, such step is rejected and time step decreased by half. 
In this way, time step is decreased when rare coarsening events occur and increased when morphology evolves slowly. 
In Figure~\ref{fig:res:ts}, we show example profile of time step size for various dimensionality considered. 
Size of time step is adjusted by few orders of magnitude.
Efficient time stepping strategy allows to perform simulations that cover several time scales. 
In the figure, we show the time scale spanning over two to three orders of magnitude when height decreases by up to 80\%.

\bibliographystyle{model1-num-names}

\end{document}